\documentclass[12pt]{article}
\usepackage[utf8]{inputenc}
\title{ Certifying Lemons}

\author{
Hershdeep Chopra
\footnote{hershdeepchopra2026@u.northwestern.edu\\ I am grateful to Jeff Ely, Marciano Siniscalchi, Bruno Strulovici, Alessandro Pavan, and Piotr Dworczak for their support and insightful discussions. All errors are my own.}
}

\date{\today}

\usepackage[T1]{fontenc}
\usepackage{lmodern}
\usepackage{mathpazo}
\usepackage[dvipsnames]{xcolor}

\usepackage{fullpage}
\usepackage{blkarray}

\usepackage{amsmath}
\usepackage{amssymb}
\usepackage{amsthm}
\usepackage{mathrsfs}
\usepackage{amsfonts}
\usepackage{cases}

\newtheorem{thm}{Theorem}
\newtheorem{lem}{Lemma}
\newtheorem{prop}{Proposition}
\newtheorem{cor}{Corollary}

\newtheorem{claim}{Claim}

\theoremstyle{definition}

\newtheorem{rem}{Remark}

\newtheorem{obs}{Observation}
\usepackage{listings}

\usepackage[capposition=top]{floatrow}
\usepackage{import}
\usepackage{comment}
\usepackage{natbib}
\usepackage{ascmac} 
\usepackage{mathtools} 
\mathtoolsset{showonlyrefs,showmanualtags} 
\usepackage{bbm}            
\usepackage[at]{easylist}
\usepackage{multicol}
\usepackage{appendix}
\usepackage{hyperref}
\hypersetup{
    colorlinks=true,
    linkcolor=blue,
    filecolor=magenta,      
    urlcolor=cyan,
}
\usepackage{enumitem}
\usepackage{amsmath}
\usepackage[normalem]{ulem}
\usepackage{caption}
\usepackage{subcaption}
\usepackage{graphicx}

\usepackage{empheq} 

\newcommand{\E}{\mathbbm{E}}

\newcommand{\R}{\mathbb{R}}

\newcommand{\argmax}{\mathop{\rm arg~max}\limits}

\DeclareMathOperator{\supp}{\mathrm{supp}}

\renewcommand{\phi}{\varphi}
\renewcommand{\epsilon}{\varepsilon}

\definecolor{S_pink}{RGB}{255, 64, 159}
\definecolor{V_Orange}{RGB}{204,85,0}
\definecolor{magenta}{RGB}{95,2,31}

\usepackage{tikz-cd}

\begin{document}

\maketitle
\begin{sloppypar}
\begin{abstract}
   This paper examines an adverse selection environment where a sender with private information (high or low ability) tries to convince a receiver of having higher ability.  Without commitment or costly signaling, market failure can occur. Certification intermediaries reduce these frictions by enabling signaling through hard information. This paper focuses on a monopolistic certifier and its impact on equilibrium welfare and certificate design.
    Key findings show that the certifier provides minimal information, pooling senders of varying abilities and leaving low rents for high type senders, which typically disadvantages the receiver. However, when precise information is demanded, the certifier screens the sender perfectly, benefiting the receiver. Thus, the monopolistic intermediary has an ambiguous effect on market efficiency.
    The results emphasize the importance of high certification standards, which drive low-ability senders out of the market. Conditions for such equilibria are characterized, showing how simple threshold strategies by the receiver induce first-best outcomes. Additionally, the relationship between the characteristics of offered certificates and welfare is identified.
\end{abstract}

\section*{Introduction}

Consider a simple adverse selection environment in which a privately informed sender (either high ability or low ability) tries to convince a receiver that he is high ability. The absence of commitment or costly signaling by the sender can lead to market failure (\cite{2b080e07-2197-39e6-8e29-cfa83db9de24}). When the sender's private information can be evaluated by a (statistical \footnote{Standards such as ANSI/ASQ Z1. 4 and ISO 2859 provide guidelines for the acceptance and sampling of products. These and many other standards essentially take the form of statistical tests.}) procedure; certification intermediaries can enable signaling between the sender and the receiver, thus reducing informational frictions and extracting a portion of the generated surplus. Such intermediaries are present in various sectors, including research laboratories, consultants, reviewers, auditors, and academic testing\footnote{ \cite{blair2011roles} discuss the potential impact of third-party certification on global commerce and discuss the need for regulation of the assurance and certification industry.  }.

This paper examines markets where a single intermediary monopolizes the sale of information. The presence of a monopolist has an ambiguous \footnote{\cite{barnett2011intermediaries} highlight the need for theoretical insights regarding certification markets and how they pose regulatory challenges. } impact on the information intermediary market. Reduced price competition enables costly signaling by the sender, but monopoly power also incentivizes the provision of less information. This occurs because less informative certificates allow the intermediary to pool senders of varying abilities. 

The paper provides insights into these tradeoffs by studying equilibrium in a sequential game among the certifier, the sender, and the receiver. In general, this game admits multiple equilibria. Therefore, I propose a novel refinement based on the receiver's inertia in reacting to observable deviations. The refinement requires that the receiver deviates from his equilibrium action only if there is a compelling reason to do so.

 The analysis indicates that the certifier has the incentive to provide a minimal amount of information, as revealing less information helps the certifier pool in low ability sender while leaving small information rents to the high ability sender. This is generally against the receiver's interests, as the receiver prefers to make decisions based on the best possible information. However, when there is a demand for precise information, the certifier perfectly screens the sender, helping the receiver achieve his desired goal. The certifier excludes the low ability sender when the required informativeness of the experiment sold is large enough, as more informative certificates are less valuable to the low ability sender, resulting in increased information rent for the high ability sender. Thus, the presence of a monopolistic intermediary has an ambiguous effect on market efficiency. 
 
These results also underscore the common wisdom of setting stringent standards. If the certification required to convince the receiver is highly informative, low ability senders exit the market. Theorem 1 and Proposition 2 characterizes the conditions under which such equilibria exist and demonstrate how a simple threshold strategy of the receiver induces first best outcome for the receiver. These equilibrium certificates take the form of \textit{compliance certificates}\footnote{For example, food allergen information use procedures like enzyme-linked immunosorbent assay (ELISA) to determine compliance to regulatory standards.}.  

Beyond threshold acceptance criteria, the receiver can have a richer acceptance criteria in equilibrium; this can increase the benefit from pooling senders. Theorem 3 characterizes all possible equilibrium outcomes in terms of a simple class of certificates that are offered in equilibrium. This shows the relationship between the equilibrium welfare and the design of optimal tests. By studying the relation of the equilibrium test structure and welfare, my model can make positive (testable) statements about information intermediary markets; a researcher can deduce facts about the efficiency of the market based on observable properties of test/certification structure. 

The rest of the paper is organized as follows: 1) In section 1, I present a simple model of mediated communication between an informed sender and an uninformed receiver, 2) In sections 2 and 3, I describe the equilibrium of the game and introduce an equilibrium refinement, 3) In section 4, I present the results and predictions of the model; my first set of results (theorem 1, proposition 1 and theorem 2) discuss efficiency of equilibrium outcomes, the second set of result (theorem 3 and proposition 2) are concerned with optimal design of certificates, 4) In section 5, I discuss the possibility of uninformative equilibrium, and 6) Finally in section 5, I conclude the paper by discussing related literature. 




\section{Model}

I study a stylized model \footnote{The closest model to mine is \cite{lizzeri1999information}, see section 6 for details. } of sender and receiver. The sender is privately informed about his ability, high or low. The receiver decides to accept or reject the sender based on verifiable results of some test (statistical experiment). The sender acquires these tests from a monopolist certifier. The certifier and receiver have a common prior $\mu$ about the sender's type. The certifier can flexibly post a menu of experiments and prices \footnote{I don't allow the pricing to be contingent on the realization of the experiment.}. In particular, the certifier (mediator) can send messages conditional on both the report of the sender and the true type. The ability to condition on the sender's true type represents the certifier's "expertise"; the certifier has some ability or technology to evaluate the agent's private information cheaply. The certifier's signals are afforded credibility through some physical, contractual, or reputation reason.

\subsection{Timing}
t=1: The monopolist certifier posts a menu of experiments and prices observable to the sender and receiver.\\
t=2: The sender privately observes his type.\\
t=3: The sender privately purchases an option from the menu offered, or decides to not get certified.\\
t=4: The sender and receiver observe the realization of the purchased experiment. If no experiment was purchased the receiver observes that the sender is not certified. \\
t=5: The receiver chooses an action in $A= \{a_h,a_l\}$.

\subsection{Sender and Receiver}
The sender's type $\theta \in \{h,l\}$ represents his private knowledge about his ability. The receiver has a prior $\mu$ for the state being $h$. The receiver chooses between actions in $A= \{a_h,a_l\}$. The receiver's utility is $\nu : A \times \{ h,l\} \rightarrow \R$ such that $ \nu(a_h,h) > \nu(a_l,h)$ and $ \nu(a_l,l) > \nu(a_h,l) $. The sender has a state independent utility $u:A\rightarrow \R$ such that $0= u(a_l) < u(a_h) =1$. The receiver's optimal choice, given his beliefs $\mu$, is $a^*(\mu) = a_h$ if $\mu \geq \pi^*$ and $a_l$ otherwise. Where $\pi^* = \frac{v(a_l,l)-v(a_h,l)}{v(a_l,l)-v(a_h,l)+v(a_h,h)-v(a_l,h)}$. I assume $0<\mu < \pi^*$; meaning that the receiver is ex-ante pessimistic about the sender's ability.

\subsection{Monopolist Certifier}
The certifier acts like a mediator between the sender and receiver. 
The certifier publicly announces a distribution of messages (signals) which depends on both the reported type of the sender and the true type. The sender is privately charged a fee based on his reported type. The certifier then publicly announces the realized message based on the true and reported type. I restrict attention to the certifier setting an upfront fee instead of a fee conditional on the realized signal\footnote{Allowing a fee conditional on the signal realization, enables the certifier to charge a fee that's correlated to the true state. The work of \cite{faure2009ownership} considers a certification model with outcome contingent fees, in particular, they show full surplus extraction by the certifier.}.

\subsection{Information}

The sender verifiably\footnote{The verifiability can be seen as a consequence of the certifier staking their reputation on the claim or some contractual restriction.} reports the realization of statistical experiment \footnote{For a topological space $X$, the set $\Delta(X)$ is the set of all Borel probability measures of $X$.} $\sigma: \{h,l\} \rightarrow \Delta([0,\infty])$. As the likelihood ratio is a sufficient statistic for binary decision problems, I identify each signal realization $e\in [0,\infty]$ for the experiment $\sigma$ with its corresponding\footnote{Note that when restricted to $[0,\infty)$ absolute continuity holds, $\sigma(.|h) \ll \sigma(.|l)$. Thus $\frac{d\sigma(x|h)}{d\sigma(x|l)}$ is well defined in this restricted set. I extend the definition to set $[0,\infty]$ by requiring $\frac{.}{0}:= \infty$.} likelihood ratio $\frac{d\sigma(e|h)}{d\sigma(e|l)}\in [0,\infty]$.
I refer to the realization $e\in [0,\infty]$ of the experiment $\sigma$ as evidence or signal realized by $\sigma$. The signal realization $e$ can be verifiably disclosed even if the experiment $\sigma$ is unknown. This allows for a tractable representation
\footnote{The restriction on the message space to $[0,\infty]$ is not essential. We can allow for any message space that is homeomorphic to $[0,\infty]$. But using $[0,\infty]$ as the message provides for a convenient representation of experiments.} of the distribution over messages (signals) that the certifier uses.
For any report $\hat{\theta}\in\{h,l\}$ and signal $e\in [0,\infty]$, the probability of realization $e$  conditional on the true type $\theta$, i.e. $\sigma_{\hat{\theta}}(e|\theta)$, is a function of $e$ and $\sigma_{\hat{\theta}}(e|h)$. Let $\Sigma$ represent the set of all experiments.

\subsection{Strategies}

The certifier posts a menu of prices and experiments $m=\{(\rho_{\theta},\sigma_{\theta})_{\theta\in \{h,l\}}, (\Phi,0)\}$. Where $\sigma_\theta:\{h,l\}\rightarrow\Delta([0,\infty])$ and $\rho_\theta\in [0,1]$. The sender has the option not to buy any of the offered experiments. Thus I require every menu to include a "no certification" \footnote{More precisely, each menu contains a deterministic experiment that takes value $\Phi$.} signal ($\Phi$) at $0$ cost. The set of all menus is $\mathcal{M}$.

The receiver's strategy is to choose an action $a\in\{a_h,a_l\}$ given some $m\in\mathcal{M}$ and $e\in ([0,\infty]\cup \Phi)$; this is given by a measurable function  $~\zeta: \mathcal{M}\times ([0,\infty] \cup \Phi)\rightarrow \{a_h,a_l\}$.
Let $E_m^{\zeta}:= \{e\in [0,\infty]\cup \Phi~|~ \zeta(m,e) = a_h\}$. Each of the sender's strategies corresponds to a collection $(E^{\zeta}_m)_{m\in \mathcal{M}}$. Whenever convenient I will omit writing $\zeta$ as part of the receiver strategy and express it in terms of collection of sets $(E_m)_{m\in \mathcal{M}}$.

The sender's strategy is a selection rule that specifies the menu item chosen by the receiver given his type $\theta$, and the posted menu $m\in \mathcal{M}$. The selection strategy is given by  $\gamma_{\theta}: \mathcal{M}\rightarrow \Sigma \cap \Phi$, i.e. $\gamma_{\theta}(m)\in \{\sigma^m_h,\sigma^m_l,\Phi\}$. The sender's optimal choice when he faces a menu $m$ and anticipates the acceptance set $E$ is given by:
\[\gamma_\theta^*(m,E) := \argmax_{\sigma'\in \{\sigma^m_h,\sigma^m_l,\Phi\}}\int_{E}d\sigma'(e|\theta) - p(\sigma')\]

Where $p(\sigma') =\rho_{\theta}$ if $\sigma'=\sigma_\theta$ and $p(\sigma') = 0$ if $\sigma'=\Phi$.

\subsection{Histories}

The certifier's history is the empty history $h_{ms}=\varnothing$. The sender moves at a history $h_s = (m,\theta)\in \mathcal{M}\times \{h,l\}$. The receiver's history is $h_r=(m,e)\in \mathcal{M}\times ([0,\infty]\cup \Phi)$.




\section{Equilibrium}

The solution concept is PBE\footnote{I focus only on pure strategies. Moreover, I require in equilibrium the receiver always breaks ties in favor of the sender and the sender breaks ties in favor of truthful reporting (meaning if $\sigma^m_{h}, \sigma^m_{l} \in \gamma_{\theta}^*(m, E_m) $ then $\gamma_{\theta}^*(m) = \sigma_{\theta}^m$).} in pure strategy with tie-breaking assumptions. An \textbf{equilibrium strategy profile} of the game is given by the tuple $((E_m)_{m\in \mathcal{M}},\gamma^*, m^*)$ which satisfies the following: 

\begin{easylist}[itemize]
    @ \textbf{Certifier rationality:} $m^*\in \argmax_{m\in\mathcal{M}} \E_{\mu}[ p(\gamma^*_{\theta}(m))] $
    @ \textbf{Sender rationality:} for all $m=((\sigma_h,\rho_h),(\sigma_l,\rho_l),\Phi)$ the sender's selection rule:
\[\gamma_{\theta}^*(m)\in\argmax_{\sigma'\in \{\sigma_h,\sigma_l,\Phi\}}\int_{E_m}d\sigma'(e|\theta) - p(\sigma')\]
In particular $\gamma^*_{\theta}(m) \in \gamma_\theta^*(m,E_m)$ for each $m\in \mathcal{M}$.
@ \textbf{Bayes rule where possible:}
    @@ At the history $(m,e)$ such that $e \in \supp(\gamma_{\theta}^*(m)(.|\theta))$ for some type $\theta\in \{h,l\}$, the receiver's belief about the sender's type being high is given by Bayes rule:
    \[\mu_{m,e} = \frac{d\gamma_h^*(m)(e|h)\mu}{d\gamma_h^*(m)(e|h)\mu+d\gamma_l^*(m)(e|l)(1-\mu)}\]
    @@ At history $(m,e)$ such that $e\not \in \supp(\gamma_{\theta}^*(m)(.|\theta))$ for both types $\theta\in \{h,l\}$. The receiver has arbitrary beliefs $\mu_{m,e}$ about the sender's type.
@ \textbf{Receiver rationality:} for any $m\in \mathcal{M}$, the receiver's acceptance set satisfies: \[E_m=\{e\in [0,\infty]\cup \Phi| a_h\in \argmax_{a\in\{a_l,a_h\}} \E_{\mu_{m,e}}[v(a,\theta)]\}\]
    
\end{easylist}
 
 I refer to $E_{m^*}$ as the receiver's equilibrium  (or on-path) \textbf{acceptance set} \footnote{A stronger version of PBE might require the element $e=\infty\in E_m$ for every $m \in \mathcal{M}$. Such a restriction could represent aspects of objectiveness in the evidence produced by the certifier. 
     This affects none of the results from section $4$ onwards, so I won't impose this restriction.}. Let $\overline{\mathcal{E}}$ represent the set of all such equilibrium. Note that given $(E_m)_{m\in \mathcal{M}}$, sequential rationality and tie-breaking uniquely pin down $\gamma^*$, thus whenever convenient I will omit explicitly mentioning the sender's equilibrium selection strategy $\gamma^*$.
     
     Define  $(E_{m^*},m^*):=((m, E'_m)_{m\in \mathcal{M}},\gamma^*,m^*)$ where $E'_m = \varnothing$ when $m\neq m^*$ and $E'_{m^*}=E_{m^*}$.

 \begin{lem}
    If $((E_m)_{m\in \mathcal{M}},m^*)\in \overline{\mathcal{E}}$ then $(E_{m^*},m^*)\in \overline{\mathcal{E}}$.
 \end{lem}
 
 \begin{proof}
     The last three conditions in the definition of equilibrium are satisfied as   
     $(( E_m)_{m\in \mathcal{M}},\gamma^*,m^*)\in \overline{\mathcal{E}}$. By definition of the off-path actions of the receiver $E_m'$, it's immediate that certifier rationality for $m^*$ also holds.
 \end{proof}

By lemma 1, we know $m$ is offered on-path in some equilibrium along with the receiver's on-path acceptance set $E_m$ if and only if
\[\frac{\int_{e}d\gamma^*_{h}(m)(e|h)}{\int_{E'}d\gamma^*_{l}(m)(E'|l)}\geq \frac{1}{l(\mu)}  ~~~~~~~~~~~\forall ~ E' \subset E_m \cap \supp(\gamma^*_l(m)(.|l))\]

 A menu $m^*\in \mathcal{M}$ is \textbf{valid} wrt $E$ if $(E,m)\in \overline{\mathcal{E}}$.
 Here $(E,m) := ((E_{m'})_{m'\in\mathcal{M}}, m)\in \overline{\mathcal{E}}$ such that $E_m = \varnothing$ whenever $m\neq m^*$ and $E_{m^*} =E$.

Essentially, the receiver can hold arbitrarily pessimistic beliefs following a deviation by the certifier. This makes the $m^*$ trivially revenue maximizing among all menus. 

\begin{rem}
An immediate conclusion from receivers ex-ante pessimism ($\mu< \pi^*$) is that for any equilibrium $((E_m)_{m\in \mathcal{M}}, \gamma^*, m^*) \in \overline{\mathcal{E}}$ it must be that $\Phi \not \in E_m$ for all $m\in \mathcal{M}$. This means that not getting certified never leads to acceptance by the receiver in any equilibrium of the game. This is easy to see, as if $\Phi \in E_m$ then at the subgame following the menu choice $m$ of the certifier either the sender or the receiver has a profitable deviation. Thus in the rest of the paper (except for section 5) all acceptance sets $E$ are assumed to be such that $\Phi\not \in E$. 
\end{rem}
Fix some set $E\subset  [0,\infty]$, the menu $m=\{(\sigma_{\theta}^m,\rho^m_\theta)_{\theta\in \{h,l\}},(\Phi,0)\} \in \mathcal{M}$ is \textbf{obedient} wrt $E$ if the menu is incentive compatible with respect to $E$ for both the sender and receiver. Meaning $m$ satisfies the following: 

\begin{easylist}
        @ \textbf{Sender IC and IR}
        \[\int_{E}d\sigma_{\theta}^{m}(e|\theta)-\rho_{\theta}\geq \int_{E}d\sigma_{\theta'}^{m}(e|\theta) - \rho_{{\theta'}} ~~~~~~~\forall ~\theta,\theta'\in \{l,h\}\]
        \[\int_{E}d\sigma_{\theta}^{m}(e|\theta)\geq \rho^m_\theta~~~~~~~~~~~~~~~~~~~~~~~~~~~~~~~~~~~~\forall ~ \theta\in\{l,h\}\]
\end{easylist}


\begin{easylist}
     @ \textbf{Receiver obedience}
        \[ \frac{\int_{E'}d\sigma_h^{m}(e|h)}{\int_{E'}d\sigma_l^{m}(e|l)}\geq \frac{1}{l(\mu)} ~~~~~~~~~~~~~~~~~~~\forall ~ E' \subset E \cap \supp(\sigma^m_l(.|l)) \]
\end{easylist}

Where $l(\mu)=\frac{\mu(1-\pi^*)}{\pi^*(1-\mu)}$. Let the certifier's revenue from an obedient menu $m$ be given by $ \text{rev}(m)= \mu\rho_h^{m} +(1-\mu)\mu\rho_l^{m}$.
For each $E\subset [0,\infty]\cap \Phi$, the set of menus obedient with respect to $E$ is given by $\mathcal{M}_E$.

Two equilibria $((E_m)_{m\in\mathcal{M}},m^*)$ and $((E'_m)_{m\in\mathcal{M}},m')$ are \textbf{outcome equivalent} if:

\begin{easylist}
 @ $\gamma^*_{\theta}(m^*)(.|h)|_{E_{m^*}}=\gamma^*_{\theta}(m')(.|h)|_{E'_{m'}} ~~\forall~~\theta~~~~~~~~~~~~~~~~~~~~~~~~$     (Experiment selection)

 @ $\E_{\mu}[ p(\gamma^*_{\theta}(m^*)] = \E_{\mu}[ p(\gamma^*_{\theta}(m')]~~~~~~~~~~~~~~~~~~~~~~~~~~~~~~~~~~~~~~$     (Revenue)

\end{easylist}
The first condition requires that in both equilibria the conditional distribution of signals restricted to the receiver's equilibrium acceptance set is the same \footnote{When the distribution of signals that lead to acceptance by the receiver is estimatable by some external researchers, the certification menu allows the researcher to infer facts about the equilibrium based on observable test structure. }. The second condition requires that the certifier earns the same revenue in both equilibria. In particular, these conditions imply that $E_{m^*} = E'_{m'}$. Moreover,  whenever the equilibrium menus $m^*,m'$ are obedient wrt $E_{m^*}$ and $E'_{m'}$ respectively then $\sigma_{\theta}^{m^*}|_{E_{m^*}} = \sigma_{\theta}^{m'}|_{E'_{m'}}$. As defined, outcome equivalence implies payoff equivalence but is not implied by payoff equivalence. If two equilibrium outcomes are equivalent then the distribution of messages conditional on the receiver accepting is the same across the equilibria for each type of sender. 

The following observation highlights the key role played by obedient menus.

 \begin{obs}
     If a menu $m^*\in \mathcal{M}$ is offered in equilibrium, then there exists an outcome equivalent equilibrium $((E')_{m\in\mathcal{M}},\gamma',m')$ such that $m'\in\mathcal{M}_{E'_{m'}}$. If a menu $m^*$ is obedient wrt some $E\subset [0,\infty]$, then an equilibrium exists in which the on-path menu is $m^*$.  
     
 \end{obs}

 \begin{proof}
   See appendix.
 \end{proof} 

 \begin{lem}
    Any equilibrium $((E_m)_{m\in\mathcal{M}},\gamma,m^*)\in \overline{\mathcal{E}}$ in which $e=0\in E_m^*$ is outcome equivalent to $((E'_m)_{m\in\mathcal{M}},\gamma'm')\in \overline{\mathcal{E}}$, where $E_m'=E_m$ for $m\neq m^*$, $E'_{m^*}=E_{m^*}\setminus\{0\}$, $m'=m^*$ and $\gamma=\gamma'$. 
\end{lem}
\begin{proof}
    Whenever $e=0\in E_{m^*}$, the signal $e=0$ has probability $0$ of being disclosed in equilibrium. Thus removing $e=0$ from $E_{m^*}$ doesn't affect the equilibrium outcomes. 
\end{proof}

Let the set of equilibrium in which the on-path menu is obedient\footnote{ and $e=0$ is not part of receiver's equilibrium strategy $E_{m^*}$.} be given by $\mathcal{E}$. As I am interested in the equilibrium outcomes it's without loss to restrict attention to equilibrium in $\mathcal{E}$. In usual mechanism design terms (\cite{forges1986approach}, \cite{myerson1986multistage}), observation 1 shows that restricting to truthful mechanisms is enough, but as outcome equivalence is stronger than payoff equivalence so restricting to a direct mechanism (pass- fail experiments) isn't necessarily without loss for the test design objective in mind. 

\section{Refinement}

Observation 1 shows that any menu that is obedient with respect to some $E$ is offered on-path in some equilibrium of the game. This conclusion relies on allowing the receiver to have arbitrary off-path beliefs. To focus on equilibria where the on-path menu is revenue maximizing with respect to the receiver's on-path acceptance set $E$, I propose a refinement of the receiver's off-path beliefs which selects for such equilibria.

The proposed refinement imposes a form of consistency on how the receiver evaluates evidence following the certifier's deviation. Essentially, the receiver does not change his acceptance set unless there is a compelling reason to do so.\footnote{The general idea of stability and inertia in deviating from equilibrium action are explored in terms of refinements in \cite{kohlberg1986strategic} and the literature citing it. The evolutionary foundation of these ideas is also discussed in \cite{young1993evolution}, and \cite{kandori1993learning}.}

\textbf{Refinement (Consistent Evaluation):} Let the set of refined equilibrium be given by:\[\overline{\mathcal{E}_r}:=\{((E_m)_{m\in\mathcal{M}},\gamma^*,m^*)\in \overline{\mathcal{E}}~| ~ \forall~m\in \mathcal{M},~(E_{m^*},m)\in \overline{\mathcal{E}} \implies E_m=E_{m^*}\}\]

Consider an equilibrium $((E_m)_{m\in \mathcal{M}},\gamma^*,m^*).$ If $m^*,~E_{m^*}$ are the on-path menu and receiver's acceptance set, then after observing a menu $m$ valid with respect to $E_{m^*}$, the receiver is sequentially rational to use acceptance set $E_{m^*}$ given that the sender' choice is in $\gamma^*(m,E_{m^*})$. Moreover, the type $\theta$ sender anticipates this and chooses $\sigma\in \gamma_{\theta}^*(m,E_{m^*})$. However, if a menu $m$ is not valid wrt $E_{m^*}$ then using the acceptance set $E_{m^*}$ is not rational for the receiver, given the senders choice is $\gamma^*(m,E_{m^*})$. Thus, the refinement requires that following the certifier's deviation; the receiver does not change his acceptance set whenever it's sequentially rational to do so, given that the sender anticipates facing the on-path acceptance set.

The proposed refinement selects for equilibria in which both the sender and receiver form consistent expectations about outcomes, even after an observable deviation in the testing structure. This means that even if the format or content of tests changes slightly, the interpretation of results remains the same. A particular consequence is that for any equilibrium that survives the refinement, the certifier can increase the prices of the offered experiments a little until he extracts second best surplus.


Let $\mathcal{M}_E^r$ be the set of all such menus that maximize the certifier's revenue among all $m\in \mathcal{M}$ that are obedient wrt $E\subset (0,\infty]$. I refer to some $m\in \mathcal{M}_E^r$ as a revenue-maximizing menu wrt $E$. By applying observation 1, we can define $\mathcal{E}_r\subset \overline{\mathcal{E}_r}$ to be the set of equilibrium such that the equilibrium menu $m^*\in \mathcal{M}_{E_{m^*}}$, where $E_{m^*}$ is the receiver equilibrium acceptance set. The following lemma the property of $\mathcal{E}_r$ eluded in last paragraph.

\begin{lem}
    If the $((E_m)_{m\in\mathcal{M}}),m^*)\in \mathcal{E}_r$ then $m^* \in \mathcal{M}_{E_{m^*}}^r.$ Moreover, if $m^* \in \mathcal{M}_E^r~$ for some $E\subset [0,\infty]$ then there exists an equilibrium in $\mathcal{E}_r$ such that $m^*$ is offered on-path and receiver's on-path acceptance set is $E$.
\end{lem}

\begin{proof}
    See appendix
\end{proof}

\begin{rem}
    If two equilibria $((E_m)_{\mathcal{M}},\gamma,m^*)$, $((E'_m)_{\mathcal{M}},\gamma',m') \in \mathcal{E}_r$ with $E_{m^*} = E'_{m'}$ are such that $\sigma_{\theta}^{m^*}$ and $\sigma_{\theta}^{m'}$ only differ on $E_{m^*}^C$, then the two equilibria are outcome equivalent.
\end{rem}

\section{Equilibrium Outcomes}

In this section, I will focus on equilibrium in $\mathcal{E}_r$ such that on-path $e=1\not \in E_{m^*}$. These equilibria constitute informative equilibria. \footnote{I elaborate on this in section 5. }

\subsection{Separating Equilibrium (Receiver Optimal)}

A consequence of modeling information provision through a certifier is the existence of equilibrium with the receiver's first best outcome. In the canonical Bayesian persuasion (\cite{kamenica2011bayesian}) setting this outcome is only achievable when the sender commits to fully informative experiments. With the presence of a certifier, the full information outcome is achievable even when the equilibrium information structure is not fully revealing. 

I call an equilibrium $((E_m)_{m\in \mathcal{M}},m^*)\in \mathcal{E}$ \textbf{separating} whenever 
\[\int_{E_{m^*}}d\sigma_h^{m^*}(e|h)=1 \text{ and } \int_{E_{m^*}}d\sigma_l^{m^*}(e|h)=0\]

\begin{thm}
If an equilibrium $((E_m)_{m\in\mathcal{M}}),m^*)\in \mathcal{E}_r$ is separating then $e^*:= \inf(E_{m^*}) \geq \frac{1}{\mu}$. Moreover, if for an equilibrium $((E_m)_{m\in\mathcal{M}}),m^*)\in \mathcal{E}_r$ it holds that $e^*:= \inf(E_{m^*}) > \frac{1}{\mu}$, then the equilibrium is separating.
\end{thm}

\begin{proof}
    See appendix
\end{proof}

    The on-path experiments in a separating equilibrium take the form of a single compliance certificate, meaning in order to be accepted the experiment needs to generate a precise signal about the sender's type. Moreover, the theorem shows the existence of separating equilibria when the on-path experiment is an imperfect quality certification ( \cite{de1991economic} and \cite{strausz2010separating}). Although the high type sender is accepted with probability 1, the low type sender is also accepted with positive probability conditional on buying the experiment.

    The theorem highlights the power of receiver commitment to high standards \footnote{ \cite{leland1979quacks} and \cite{demarzo2019test} have a similar message in different certification environments.}. If the receiver can commit to only accepting evidence that is above a certain threshold likelihood ratio, then the receiver can guarantee himself the first best outcome. The certifier acts as a screening device to screen the sender for the receiver. When the receiver can insist on high standards, he induces equilibrium in which the certifier offers contracts that exclude the low type and pass the high type with probability 1. \\
    The certifier excludes the low type for the revenue-maximizing motive. By requiring a high threshold likelihood ratio, the sender increases the informativeness of experiments offered by the certifier. This higher informativeness then leads to the exclusion of the low type as including the low type would require leaving large rents to the high type.



\subsection{Naive Receiver}

In this section, I focus on how a naive receiver affects the equilibrium. Formally, a \textbf{naive receiver} is described by the strategy $E_m = [\frac{1}{l(\mu)},\infty]$ for all $m$. The set of all equilibrium in $\mathcal{E}$ for which the receiver is naive is given by $\mathcal{E}_n$.

\begin{prop}
    The set $\mathcal{E}_n\subset \mathcal{E}_r$. If $((E_m)_{m\in\mathcal{M}},m^*)\in \mathcal{E}_n$ then $\sigma_h^{m^*}=\sigma_l^{m^*}$ and $\sigma_h^{m^*}\left(\frac{1}{l(\mu)}|h\right)=1$. Moreover, $((E_m)_{m\in\mathcal{M}},m^*)\in \mathcal{E}_n$ is separating (if and) only if $\mu (<)\leq 2-\frac{1}{\pi^*}$. 
\end{prop}

\begin{proof}
 See appendix  
\end{proof}

   When the receiver is naive as described above, Proposition 3 states that the market price discriminates, i.e. only sells to high type buyers, when the fraction of high type buyers is small enough. Although the smaller market fraction reduces the benefit from price discrimination, the demand for more precise signals increases when the prior beliefs are low, this in turn increases the gain from extracting surplus from the high type. Price discrimination then depends on the interplay between these forces; proposition 1 characterizes conditions under which the increase in demand for more precise signals outweighs the disincentive from smaller market size. By adopting a naive approach\footnote{Behaviourally this is a receiver who forms beliefs by considering signals at face value. For example after observing a realization $e\in [0,\infty]$, the naive receiver updates his belief about the state as following $Pr(\theta=h|e) = \frac{e\mu}{e\mu + 1-\mu}$.} to evaluating evidence, the receiver can guarantee himself first best when he is sufficiently pessimistic about the sender's prior ability.

\subsection{Sender Optimal}

In this section, I will characterize the sender optimal equilibrium in the set $\mathcal{E}_r$. 

\begin{thm}
    The equilibrium $((E)_{m\in\mathcal{M}},m^*)\in \mathcal{E}_r$ achieves the highest ex-ante expected payoff among all equilibrium in $\mathcal{E}_r$ only if $\min(E_{m^*})=\frac{1}{\mu}$. Moreover, the equilibrium menu $m^*$ is such that $\sigma_h^{m^*}\left(\frac{1}{\mu}~|~h\right) = 1.$
\end{thm}

\begin{proof}
   See appendix
\end{proof}

    The total surplus of trade (sum of the certifier and sender surplus) is maximized in the \textbf{KG equilibrium}. Where both types get KG menu\footnote{A menu such that $\sigma_h=\sigma_l$ and $\sigma_h(\frac{1}{l(\mu)}|h)= 1$. This is the optimal experiment when the sender has commitment power \cite{kamenica2011bayesian}.} at no cost and the receiver accepts if he observes $e = \frac{1}{l(\mu)}$. 
    
    When the prior $\mu > 2-\frac{1}{\pi^*}$, the experiment offered in the sender optimal equilibrium is more informative than the KG menu. This equilibrium has a total surplus from trade strictly less than the maximum possible surplus.  When the prior $\mu < 2-\frac{1}{\pi^*}$, the experiments offered in the sender optimal equilibrium is less informative than the KG experiment. Moreover, this equilibrium also maximizes the total surplus from trade. \\ 
    In particular, when the receiver is sufficiently confident about the sender's ability (but not enough to outright accept him), the optimal sender equilibrium leads to an inefficient trade surplus. The inefficiency in trade surplus is due to the increased informativeness of equilibrium experiment; the greater informativeness helps high type sender to earn a greater share of the surplus but decreases the total surplus. 

\subsection{Disclosure Without Screening}

In this section I will examine the case when the certifier can not partake in monopolistic screening; the certifier is restricted to offer a single experiment and price pair. \\
When a single menu item is offered, proof of theorem 1 and proposition 1 show that the equilibrium acceptance set $E_{m^*}$ is such that $\inf(E_{m^*}) \geq \frac{1}{l(\mu)}$. In particular, if $\mu < 2 - \frac{1}{\pi^*}$ then all equilibrium in $\mathcal{E}_r$ are separating equilibrium (see section 4.1). In the absence of the certifier's ability to price discriminate, the receiver achieves the first best outcome whenever he is sufficiently pessimistic. When $\mu \geq 2-\frac{1}{\pi^*}$, there exist equilibrium pooling. Thus whenever the adverse selection problem is severe $(\mu < 2-\frac{1}{\pi^*})$, then the receiver is always (weakly) better off when the monopolist can not screen. 

\subsection{Structure of Equilibrium Certificates}

This section is primarily technical, it establishes that the equilibrium in $\mathcal{E}_r$ can be studied by solving a constrained linear optimization problem. I characterize all possible equilibrium outcomes in $\mathcal{E}_r$ by first characterizing $\mathcal{M}^r_E$ in terms of a simplified linear optimization problem. Then using this, I focus on a simpler set of menus that generate all feasible menus in the simplified problem. Finally, I find the optimal menus among these simple menus. In particular, this shows that the model studied in this paper affords a tractable analysis of the information intermediary market.


\begin{prop}
    Fix some $E\subset (0,\infty]$. Let $m^*= ((\sigma_{\theta}^{m^*},\rho_{\theta}^{m^*})_{\theta\in \{l,h\}})$ be obedient wrt $E$. Then $m^*\in \mathcal{M}_E^r$ if and only if $m^*$ is such that 
    \\$\rho^{m^*}_h= \int_E\left[d\sigma^{m^*}_h(e|h)-\left(1-\frac{1}{e}\right)d\sigma^{m^*}_l(e|h)\right]$, $\rho^{m^*}_l = \int_Ed\sigma^{m^*}_l(e|l)$ and solves the following optimization problem:
   \[ \max_{(\sigma_h^m(.|h),\sigma_l^m(.|h))\in \Delta([0,\infty])\times \Delta([0,\infty])} ~~\mu\int_E d\sigma_h^m(e|h) + \int_E\left( \frac{1}{e}-\mu\right) d\sigma_l^m(e|h)\]
   subject to
   \[ \int_E \left(1-\frac{1}{e}\right)d\sigma_h^m(e|h) \geq \int_E \left(1-\frac{1}{e}\right)d\sigma_l^m(e|h)\geq 0\]
   \[\frac{\int_{E'}d\sigma_h^m(e|h)}{\int_{E'}d\sigma_l^m(e|l)}\geq \frac{1}{l(\mu)}~~\text{for all } E'\subset E\cup \supp(\sigma_l(.|l))\]
\end{prop}
     \begin{proof}See appendix
     \end{proof}


\begin{rem}
    The proposition follows by noting two properties of the optimal menu. First, it leaves zero rent to the low type. Second, the high type's willingness to pay for either of the offered experiments is weakly greater than the willingness to pay of the low type sender. 
\end{rem}

\begin{cor}
  If $m\in \mathcal{M}^r_E$ and $\text{ rev}(m)>0$ for some $E\subset [0,\infty]$ then $\int_E d\sigma^m_h(e|h)=1$.
\end{cor}
\begin{proof}
   See appendix
\end{proof}

    This shows that any equilibrium menu in $\mathcal{E}_r$ leads to the high type being accepted with probability 1. In particular, the receiver faces no distortion whenever the sender is high type. But as shown by theorem 1, in general, the receiver's payoff might be distorted when the sender is low type. 
    Whenever $m\in \mathcal{M}_E$ for some $E$, information rent is only conceded to the high type. This rent is given by,  $~\text{rent}(m) = \int_E\left(1-\frac{1}{e}\right)d\sigma_l^m(e|h)$.

\subsubsection{Equilibrium Characterization}

In this section, I restrict attention to equilibrium with countable acceptance sets\footnote{I prove the statement in the appendix for discreet support distribution.}.  

For any $E\subset [0,\infty]$, let \[\mathcal{T}(E):= \{m\in \mathcal{M}^r_E| ~|\supp \sigma_h^m(.|h)\cap E|\leq 3,~~| \supp \sigma_l^m(.|h)\cap E| \leq 2\}\]

\begin{thm}
    If $((E_m)_{m\in\mathcal{M}}, m^*)\in \mathcal{E}_r$ such that $E_{m^*}$ is countable, then there exists an outcome equivalent equilibrium  $((E'_m)_{m\in\mathcal{M}}, m')\in \mathcal{E}_r$ such that $m'\in \textbf{cvx}(\mathcal{T}(E_{m^*}))$.
\end{thm}

\begin{proof}
    See appendix
\end{proof}

    The proof of the theorem involves solving the linear optimization problem in proposition 2, to do so, I first deal with the point-wise inequalities $\frac{\sigma_h(e|h)}{\sigma_l(e|l)} \geq \frac{1}{l(\mu)}$ for all $e\in E\cap \supp(\sigma_l(e|l))$. Then I proceed by solving the simplified problem by finding the extreme points of the feasible set of experiments\footnote{See appendix for details}. 
    The theorem establishes that all equilibrium outcomes in $\mathcal{E}_r$ are generated by a simple class of menus. In particular, it characterizes the implementable rent distributions for equilibrium in $\mathcal{E}_r$.

    Moreover, along with lemma 6 and 7 it establishes existence equilibrium with the following properties:\\
    \textbf{Partial pooling} equilibrium in which $\supp(\eta_l)\cap E \subsetneq \supp(\eta_h)\cap E$. (Lemma 7)\\
    \textbf{Bad news} equilibrium in which $\exists~ e\in E$ such that $e<1$ and $\eta_l(e) > 0$. (Lemma 6)
    
    Partial pooling equilibrium corresponds to the optimal menu offered by the certifier that consists of experiments with outcomes that only a high type can achieve with positive probability. Essentially the richness of the receiver's on-path acceptance set allows the certifier more possibilities to pool senders of varying abilities.

\section{(Un)informative Equilibrium}

From the discussion in section 4, we see two important assumptions that lead to informative equilibrium outcomes: 1) An uninformative signal $(e=1)$ is never accepted on-path, and 2) The receiver holds a pessimistic prior $\mu< \pi^*$. As I will show below, the first of these assumptions avoids pathological outcomes, however, the second assumption has more substantive implications. 

\subsection{Certifier Optimal}

When the receiver is pessimistic, this is $\mu < \pi^*$, then $\mathcal{E}_r$ has a pathological equilibrium in which the signal $e=1$ is part of the receiver's on-path acceptance set. \\
There is an equilibrium in $\mathcal{E}_r$ in which $e=1$ is accepted by the receiver on-path, moreover, the on-path menu $m^*$ is such that 
\[ \sigma_h^{m^*}(1|h) = 1,~ \rho_h^{m^*} =1,~ \sigma_l^{m^*}(1|h) = l(\mu),~ \rho_l^{m^*} = l(\mu)\]
This outcome\footnote{The equilibrium outcome mentioned in section 5.1 is informative in the sense that it leads to a receiver posterior that is different than his prior. However, the informativeness of realized signals is purely through self-selection, whereas the outcomes in section 4 were informative through both self-selection and dependence on the true type.}  is "uninformative" as the certifier can achieve this outcome as a result of state-independent garbling; assuming truthful reporting, the certifier will fully disclose whenever the reported type is high and choose a noisy disclosure (garbel the report) whenever the reported type is low. Thus the outcome doesn't depend on the intermediary's testing abilities. This is in contrast with many real-world certification environments, where the certifier represents experts who conduct verifiable and informative tests.

\subsection{Optimistic Receiver}

When the receiver is optimistic $(\mu \geq \pi^*)$ the set $\mathcal{E}_r$ admits only two equilibrium outcomes. First, the outcome in which the on-path acceptance set of the receiver contains $\Phi$. In such an equilibrium, all types of senders are expected not to get certified. Thus both types of senders choose $\Phi$ on-path and the receiver takes an action based on his prior belief.

In the second outcome, the certifier offers a single uninformative experiment at price $1$; the on-path menu $m^*$ is such that 
\[\sigma^{m^*}_h(1|h)= \sigma^{m^*}_l(1|h),~ \rho^{m^*}_h=\rho^{m^*}_l =1\]
Moreover, in this equilibrium, the receiver interprets no certification $(\Phi)$ as a signal coming from the low type sender. Thus on-path, both types of senders will choose to purchase the uninformative experiment. 
This outcome is similar to the parasitic intermediary result in \cite{lizzeri1999information}, I discuss this further in the next section.

\section{Literature Review and Discussion}

The issue of signaling and market efficiency in pure adverse selection has been widely studied in economic literature. The seminal paper by\cite{2b080e07-2197-39e6-8e29-cfa83db9de24} shows how markets can unravel in the presence of adverse selection, \cite{viscusi1978note} demonstrates that quality certification can provide an alternative to exiting the market for high type producers. \cite{spence1978job} shows that undertaking costly actions can help signal private information in the context of labor markets. 

This paper, like \cite{lizzeri1999information}, considers a monopolistic information intermediary, which can produce hard (verifiable) information about the sender's private type. Unlike models of signaling where the costly action undertaken for signaling is wasteful, here the costly action directly corresponds to the payoff of the intermediary. I recover some of the results from \cite{lizzeri1999information} when the receiver in my model is optimistic.\\
However, the parasitic intermediary result no longer holds when the receiver is pessimistic; the main reason is that the pessimistic prior beliefs restrict the certifier from selling arbitrarily uninformative experiments leading to richer equilibria.
I consider both the information design and monopolist screening problem that the intermediary faces, \cite{lizzeri1999information} only focuses on the design aspect. As mentioned in section 4.4 and section 5.2, the assumption of not allowing the certifier to screen has bite in terms of equilibrium outcomes whenever the receiver is pessimistic ($\mu < \pi^*)$. 

The disclosure of hard evidence is often studied in the context of voluntary disclosure \cite{grossman1981informational}, \cite{grossman1983implicit}, \cite{milgrom1981good}, \cite{dye1985strategic}. In the context of this paper, the main takeaway from models of voluntary disclosure is the minimum principle (\cite{guttman2014not}, \cite{demarzo2019test}): non-disclosure is treated in the most pessimistic way by the receiver. The minimum principle holds in my model, as the prior belief ($\mu < \pi^*$) prescribes choosing $a_l$ whenever the receiver observes no certification.  

Although I focus on a single monopolist certifier, the analysis in the paper describes all possible equilibrium welfare. This describes outcomes when the players have varying levels of market power which reduces the share of surplus that the monopolist can capture.  

Previous works on monopolistic certification have studied the welfare implications in isolation. My work emphasizes the relationship between test design and welfare aspects in these environments. Notably, \cite{weksler2023certification} considers the test design and welfare in markets with monopolistic certifiers. However, unlike this paper \cite{weksler2023certification} does not study a monopolistic screening problem by only allowing the certifier to post a single experiment.
The paper by \cite{dasgupta2022hard} considers a general test design and screening problem similar to mine, but they focus on an environment where an uninformed sender can flexibly design a test before interacting with a receiver (first-party certification). The issue of test design is also studied by \cite{ali2022sell}, their analysis differs from mine as they are primarily concerned with outcomes when the intermediary worries about the worst-case revenue across all equilibrium outcomes. 

Closely related to this paper is \cite{corrao2023mediation}, which considers a similar monopolistic certification problem but focuses on soft information. In \cite{corrao2023mediation}, the intermediary cannot condition the experiment's outcomes on the true type of the sender and must rely solely on the reported type. They show that the mediator when restricted to soft information, can induce any receiver's belief that is consistent with hard information. This result hinges on the mediator's ability to screen senders of varying abilities; however, when the sender's type is binary, this separation in types is not achievable with soft information alone. In my model, the only equilibrium outcome implementable by soft information is the certifier optimal outcome (section 5.1).

Following the work of \cite{kleiner2021extreme}, it's common in economic theory to study outcomes of mechanism design and information design problems geometrically. The technical result of this paper follows this theme by first reducing the screening and design to a constrained linear optimization problem, then characterizing the equilibrium outcomes in terms of the extreme points of the feasible set of experiments offered by the certifier. 

The literature on monopolistic certification also considers a moral hazard environment; \cite{shapiro1986investment}, \cite{albano2001strategic}, \cite{zapechelnyuk2020optimal}. However, similar to the certification literature with pure adverse selection, these papers are silent about the design of optimal tests. This presents a future direction for extending the methodology of this paper to other certification environments where the investment in costly action is not only corresponds to the surplus of the certifier but also has a role in capital development for the receiver.
\newpage

\bibliography{ref}
\bibliographystyle{aer}

\newpage
\section{Appendix}

\subsection{Useful lemmas}

\begin{lem}
    There exists $e>1\in E$ if and only if there exists a menu $m\in \mathcal{M}_E$ such that $\text{rev}(m)>0$ (or equivalently $\int_{E}d\sigma_h^{m}(e|h)>0$).
  
\end{lem}
\begin{proof}
    Sufficiency: when $e>1\in E$, the menu $m$ such that $\sigma_h^m(e|h)=1$,  $\rho_h^m=1$ and $\sigma^m_l=\Phi, ~\rho_l^m=0$ is obedient wrt $E$.\\
    Necessity: first note that incentive compatibility implies the monotonicity of allocation:
    \[\int_{E}[d\sigma_h^{m}(e|h)-d\sigma_l^{m}(e|h)] \geq \int_{E}\frac{1}{e}[d\sigma_h^{m}(e|h)-d\sigma_l^{m}(e|h)]\]
    If $e\in E \implies e<1$, we see that only the menu $m$ with $\int_{E}d\sigma_h^{m}(e|h) = \int_{E}d\sigma_l^{m}(e|h)$ satisfy this. In particular, $\rho_h^m=\rho_l^m $. By obedience condition of the receiver we get that $\int_E d\sigma_h^m(e|h)\geq \int_E\frac{1}{el(\mu)}d\sigma_l^m(e|h)$. But this implies $\int_Ed\sigma_l^m(e|h)\geq \int_E\frac{1}{el(\mu)}d\sigma_l^m(e|h)$. As $l(\mu)<1$ and by assumption $e<1$ for all $e\in E$, this implies  $\int_{E}d\sigma_h^{m}(e|h) = \int_{E}d\sigma_l^{m}(e|h)=0$. By individual rationality of the sender, we get $\rho_h^m=\rho_l^m =0$
\end{proof}

\begin{cor}
    Let $m\in \mathcal{M}_E$ for some $E\subset [0,\infty]$, if there exists $E'\subset E\cap [0,1]$ such that $\int_{E'}d\sigma^m_h(e|h)>0$ then there exists some $E''\subset E \cap [1,\infty]$ such that $e>1$ and $\int_{E''}d\sigma^m_h(e|h)>0$.
\end{cor}

\begin{proof}
    Let there exists $E'\subset E\cap [0,1]$ such that $\int_{E'}d\sigma^m_h(e|h)>0$, if there doesn't exist $ E''\subset [0,\infty]$ with the properties mentioned above then $\int_{E}d\sigma_h^{m}(e|h)=0$. This follows from reasoning similar to the necessity part of lemma 1; to avoid violation of monotonicity of allocation $\int_{E}d\sigma_h^{m}(e|h)=\int_{E}d\sigma_l^{m}(e|h)$. Then for the menu $m$ to be obedient wrt $E$ it must hold that $\int_{E}d\sigma_h^{m}(e|h) = \int_{E}d\sigma_l^{m}(e|h)=0$. Thus the corollary holds by contradiction.\end{proof}

\subsection{Proof of Observation 1}
\textbf{Observation 1:}

     If a menu $m^*\in \mathcal{M}$ is offered in equilibrium, then there exists an outcome equivalent equilibrium $((E')_{m\in\mathcal{M}},\gamma',m')$ such that $m'\in\mathcal{M}_{E'_{m'}}$. If a menu $m^*$ is obedient wrt some $E\subset [0,\infty]$, then an equilibrium exists in which the certifier offers $m^*$ on-path.

 \begin{proof}
    If a menu $m^*$ is obedient wrt $E$, then $\gamma_\theta^*(m^*,E) = \sigma_\theta^{m^*}$ and $\mu_{m^*,e} \geq \pi^*$ for all $e\in E$. \footnote{$(E,m^*)$ is defined in lemma 1 in the main text.}These two then imply that $(E,m^*)$ is in $\overline{\mathcal{E}}$. Pessimistic off-beliefs of the receiver imply that the certifier has no incentive to deviate. The sender is best responding as the menu $m^*$ and the receiver's strategy $E$, as $m^*$ is obedient wrt E. Obedience implies that the receiver is best responding to Bayesian beliefs on-path given the equilibrium menu and the sender's selection rule. Thus $(E,m^*) \in \overline{\mathcal{E}}$.  \\
    Let $((E_m)_{m\in \mathcal{M}},\gamma^*,m^*)\in \overline{\mathcal{E}}$. If $m^*$ satisfies sender IC wrt $E_{m^*}$ this is $\gamma_\theta^*(m,E) = \sigma_\theta^{m^*}$, then by the definition of equilibrium we get that $m^*$ satisfies receiver obedience wrt $E_{m^*}$. 
    Assume that $m^*$ doesn't satisfy sender IC wrt $E_{m^*}$. This implies that for some $\theta\in\{h,l\}$ the senders selection $\gamma_\theta(m^*,E^*)\neq \sigma^{m^*}_\theta$. In this case the menu $m' := ((\gamma_\theta(m^*),p(\gamma_\theta(m^*)))_{\theta\in\{h,l\}}$ is obedient wrt $E_{m^*}$. Thus $(E_{m^*},m')\in \overline{\mathcal{E}}$ by the first part of the proof. Moreover, the equilibrium $(E_{m^*},m')$ is outcome equivalent to $((E_m)_{m\in \mathcal{M}},\gamma,m^*)$ by construction.
 \end{proof} 

    \subsection{Proof of Lemma 3}

    \textbf{Lemma 3:}\\
    If the $((E_m)_{m\in\mathcal{M}}),m^*)\in \mathcal{E}_r$ then $m^* \in \mathcal{M}_{E}^r.$ Moreover, if $m^* \in \mathcal{M}_E^r.$ for some $E\subset [0,\infty]$ then there exists an equilibrium in $\mathcal{E}_r$ for which $m^*$ is offered on-path and the on-path acceptance set is $E_{m^*}=E$.

    \begin{proof}
        If $((E_m)_{m\in\mathcal{M}},m^*)\in \mathcal{E}_r$ then by definition $((E_m)_{m\in\mathcal{M}},m^*)\in \mathcal{E}$ and $E_m=E_{m^*}$ for all $m\in \mathcal{M}_{E_{m^*}}$. Recall that if $m^*$ is offered on-path then $m^*$ maximizes the certifier's revenue among all menus $m\in \mathcal{M}$ given the sender's best responds to the receiver's strategy $(E_m)_{m\in\mathcal{M}}$. As $E_m=E_{m^*}$ for all $m\in \mathcal{M}_{E_{m^*}}$, we get that $\text{rev}(m^*)\geq \text{rev}(m)$ for all $m\in \mathcal{M}_{E_{m^*}}$. Thus $m^*\in \mathcal{M}_{E_{m^*}}^r$.\\
        For the other direction note that if $m^*\in \mathcal{M}_{E}^r$ for some $E$ then $(E, m^*)\in \mathcal{E}$. We construct the following equilibrium profile $((E_m)_{m\in\mathcal{M}}, m^*)$ where $E_m=\varnothing$ whenever $m\not \in \mathcal{M}_E$ and $E_m=E$ whenever $m\in \mathcal{M}_E$. Now note that  $((E_m)_{m\in\mathcal{M}}, m^*)\in \mathcal{E}$ as $m^*$ is obedient wrt $E_{m^*}=E$, and by construction $m^*$ maximizes the certifier's revenue given the sender's best response to the receiver's strategy. Thus $((E_m)_{m\in\mathcal{M}}, m^*)\in \mathcal{E}_r$.
    \end{proof}

\textbf{For the rest of the proofs, I restrict attention to acceptance sets $E$ such that $e=1\not \in E$. }

    \subsection{Proof of Proposition 2}

\textbf{Proposition 2:}\\
       Fix some $E\subset (0,\infty]$. Let $m^*= ((\sigma_{\theta}^{m^*},\rho_{\theta}^{m^*})_{\theta\in \{l,h\}})$ be obedient wrt $E$. Then $m^*\in \mathcal{M}_E^r$ if and only if $m^*$ is such that 
    \\$\rho^{m^*}_h= \int_E\left[d\sigma^{m^*}_h(e|h)-\left(1-\frac{1}{e}\right)d\sigma^{m^*}_l(e|h)\right]$, $\rho^{m^*}_l = \int_Ed\sigma^{m^*}_l(e|l)$ and solves the following optimization problem:
   \[ \max_{(\sigma_h^m(.|h),\sigma_l^m(.|h))\in \Delta([0,\infty])\times \Delta([0,\infty])} ~~\mu\int_E d\sigma_h^m(e|h) + \int_E\left( \frac{1}{e}-\mu\right) d\sigma_l^m(e|h)\]
   subject to
   \[ \int_E \left(1-\frac{1}{e}\right)d\sigma_h^m(e|h) \geq \int_E \left(1-\frac{1}{e}\right)d\sigma_l^m(e|h)\geq 0\]
   \[\frac{\int_{E'}d\sigma_h^m(e|h)}{\int_{E'}d\sigma_l^m(e|l)}\geq \frac{1}{l(\mu)}~~\text{for all } E'\subset E\cup \supp(\sigma_l(.|l))\]

     \begin{proof}
         To prove the proposition I will first show that the low type has zero surplus in any $m\in \mathcal{M}_E^r$. The certifier's optimization problem is given by: 
         \[ \max_{m\in M} ~~\mu\rho^m_h + (1-\mu)\rho^m_l\]
   such that 
    \[\int_{E}[d\sigma_h^{m}(e|h)-d\sigma_l^{m}(e|h)] \geq \rho^m_h-\rho^m_l \geq \int_{E}\frac{1}{e}[d\sigma_h^{m}(e|h)-d\sigma_l^{m}(e|h)]\]
    \[\int_{E}d\sigma_h^{m}(e|h)\geq \rho^m_h~~\text{ and }~ \int_{E}\frac{1}{e}d\sigma_l^{m}(e|h)\geq\rho^m_l\]
\[\frac{\int_{E'}d\sigma_h^m(e|h)}{\int_{E'}d\sigma_l^m(e|l)}\geq \frac{1}{l(\mu)}~~\text{for all } E'\subset E\]

     For revenue maximization, either low type or high type IR constraint needs to be binding. When the low type's IR constraint is binding the optimization problem reduces to the one in the proposition. This follows from substituting $\rho_l^m = \int_E\frac{1}{e}d\sigma^m_l(e|h)$
    and $ \rho_h^m = \int_Ed\sigma^m_h(e|h)$ - $\int_E\left(1-\frac{1}{e}\right)d\sigma^m_l(e|h)$. 
    \\
    Thus, to prove the proposition, we first show that the low type never earns rent in a revenue-maximizing menu. Consider for contradiction a revenue-maximizing menu $\chi \in \mathcal{M}$ that offers the low type positive rent, then  $\rho_l^\chi = \int_E\frac{1}{e}d\sigma^{\chi}_l(e|h) - \max\{\int_E\left(\frac{1}{e}-1\right)d\sigma^{\chi}_h(e|h),0\}$
    and $ \rho_h^{\chi} = \int_Ed\sigma^{\chi}_h(e|h)$. In particular ${\chi}$ is a solution to 
     \[ \max_{(\sigma_h^m(.|h),\sigma_l^m(.|h))\in \Delta([0,\infty])\times \Delta([0,\infty])} ~~(1-\mu)\int_E \frac{1}{e}d\sigma_l^m(e|h) + \int_E\left( 1-\frac{1-\mu}{e}\right) d\sigma_h^m(e|h)\]
   such that 
   \[
   \int_E \left(1-\frac{1}{e}\right)d\sigma_h^m(e|h) \geq \int_E \left(1-\frac{1}{e}\right)d\sigma_l^m(e|h)\]
   \[
   0 \geq \int_E \left(1-\frac{1}{e}\right)d\sigma_l^m(e|h)\]
    \[\frac{\int_{E'}d\sigma_h^m(e|h)}{\int_{E'}d\sigma_l^m(e|l)}\geq \frac{1}{l(\mu)}~~\text{for all } E'\subset E\]
   When the low type earns positive rent, i.e. $\int_E\left(\frac{1}{e}-1\right)d\sigma^{\chi}_h(e|h)>0$, it must be that:
   $\int_Ed\sigma^{\chi}_l(e|h) = 1- \epsilon$ for some $\epsilon>0$.  Moreover, there must exist
   some $E_l\subset E$ such that $e\in E_l\subset E\cap [0,1]$, $E_l$ is closed and $\int_{E_l}d\sigma^{\chi}_l(e|h)>0$. As $\frac{\sigma^{\chi}_h(e|h)}{\sigma^{\chi}_l(e|h)}\geq \frac{1}{el(\mu)}$ for all $e\in E$ we have that $\int_{E_l}d\sigma^{\chi}_h(e|h)>0$. Then by corollary 2 we get that there must be some $E_h \subset E\cap [0,1]$ such that $\int_{E_h}d\sigma^{\chi}_h(e|h)>0$. 
   Fix some arbitrary $e_h\in E_h$ and define $e_l:=\min(E_l)$.\\
   Note that $\int_E\frac{1}{e}d\sigma^{\chi}_h(e|h)=1$ when the rent to low type is strictly positive, as otherwise, the certifier can increase revenue by slightly increasing the probability of $e_h$ in high type's menu option. In particular if $\int_E\sigma^{\chi}_h(e|h) < 1$ and $\int_E \frac{1}{e}d\sigma^{\chi}_h(e|h) > \int_Ed\sigma_h^{\chi}(e|h)$ then $\int_{E^c}d\sigma_h^{\chi}(e|h) > \int_{E^c}d\sigma_h^{\chi}(e|l)>0$; then the certifier can improve his payoff by offering a menu with experiments $(\sigma'_h,\sigma'_l)$. Where $(\sigma'_h,\sigma'_l)$ is such that $\sigma'_l = \sigma^{\chi}_l$, and for some $e_h\in E\cap [1,\infty]$ and $\kappa >0$ small we have $\sigma'_h(e_h|h) = \sigma^{\chi}_h(e_h|h) + \kappa \min\{e_h\int_{E^c}\sigma^{\chi}_h(e|l),~ \int_{E^c}\sigma^{\chi}_h(e|h)\}$, $\sigma'_h(\infty|h) = \sigma^{\chi}_h(\infty|h) + \kappa \max\{[\int_{E^c}\sigma^{\chi}_h(e|h)-e_h\int_{E^c}\sigma^{\chi}_h(e|l)],~0\}$, and $\sigma'_h(e|h) = \sigma_h^{\chi}(e|h)$ for all $e \in E \setminus\{e_h,\infty\}$.
   

   Now to show low type can not earn positive rent construct $\sigma_\theta':\{l,h\}\rightarrow \Delta[0,\infty]$ such that $\sigma_\theta'(e|\theta') = \sigma^{\chi}_\theta(e|\theta')$ for all $\theta, \theta'$ and $e\in E\setminus(E_l\cup\{e_h\})$. 
   Moreover, set $\sigma_h'(e_h|h) = \sigma^{\chi}_h(e_h|h) + \int_{E_l}\frac{\delta}{el(\mu)}d\sigma^{\chi}_l(e|h)$, $d\sigma_h'(e|h) = d\sigma^{\chi}_h(e|h) - \frac{\delta}{el(\mu)}d\sigma_l^{\chi}(e|h)$,  $\sigma_l'(e_h|h) = \sigma^{\chi}_l(e_h|h) + \delta\int_{E_l}d\sigma_l^{\chi}(e|h)$ and $\sigma_l'(e|h) = (1-\delta)\sigma^{\chi}_l(e|h)$ whenever $e\in E_l$. Let $\rho'_h= \rho_h^{\chi}$ and $\rho'_l=\rho_l^\chi- \delta\int_{E_l}\left(\frac{1}{el(\mu)} - 1\right)\left(\frac{1}{e}-\frac{1}{e_h}\right)d\sigma_l^{\chi}(e|h) $. Define $\chi':=((\sigma'_h,\rho_h'),(\sigma'_l,\rho_l'))$. 
   Choose $\delta > 0$ small enough such that $\int_E\left( 1- \frac{1}{e}\right) d\sigma'_l(e|h) \geq \int_E\left(1-\frac{1}{e}\right) d\sigma_l'(e|h)$ and $0\geq \int_E\left( 1-\frac{1}{e}\right) d\sigma_l'(e|h)$.
   By construction, we also get that 
   \[\text{rev}(\chi')-\text{rev}(\chi)= (1-\mu)\delta\int_{E_l}\left(\frac{1}{el(\mu)} - 1\right)\left(\frac{1}{e}-\frac{1}{e_h}\right)d\sigma_l^{\chi}(e|h)>0\]
   This concludes the claim that the low type doesn't earn positive rents. \\
   The proof then follows from noting that $\int_E\left( 1-\frac{1}{e}\right)d\sigma_h(e|h)\geq 0 \implies \int_E\left( 1-\frac{1}{e}\right)d\sigma_l(e|h)\geq 0$.
   Consider for contradiction that $\int_E\left( 1-\frac{1}{e}\right)d\sigma_h(e|h)\geq 0 $ and $\int_E\left( 1-\frac{1}{e}\right)d\sigma_l(e|h)<0$. Then the constraint $\frac{\int_{E'}d\sigma_h(e|h)}{\int_{E'}d\sigma_l(e|l)}\geq \frac{1}{l(\mu)}$ is not biding for some $E'\subset E$. If instead, the constraint $\frac{d\sigma_h(e|h)}{d\sigma_l(e|l)}\geq \frac{1}{l(\mu)}$ is binding for all $e\in E$ the following calculation gives us a contradiction:
   \[ \int_E\left( \frac{1}{e}-1\right)d\sigma_l(e|h)>0\]
   \[ \implies \int_E\left( 1-e\right)d\sigma_h(e|h)>0\]
   \[ \implies \int_{E\cap\{e>1\}}\left( \frac{1}{e}-1\right)d\sigma_h(e|h) + \int_{E\cap\{e<1\}}\left( 1-e\right)d\sigma_h(e|h) >0\]
   But 
   \[\int_E\left( 1-\frac{1}{e}\right)d\sigma_h(e|h)\geq 0\]
   \[\implies \int_{E\cap\{e>1\}}\left( \frac{1}{e}-1\right)d\sigma_h(e|h) + \int_{E\cap\{e<1\}}\left( \frac{1-e}{e}\right)d\sigma_h(e|h) \leq 0\]
    \[ \implies \int_{E\cap\{e>1\}}\left( \frac{1}{e}-1\right)d\sigma_h(e|h) + \int_{E\cap\{e<1\}}\left( 1-e\right)d\sigma_h(e|h) \leq 0\]

    Finally, the claim follows from the fact that if both the following statements are true then the certifier has a profitable deviation.
    \[\exists e'\in E \text{ such that } \frac{d\sigma_h(e'|h)}{d\sigma_l(e'|h)}> \frac{1}{e'l(\mu)}\]
    and 
    \[\int_E\left( 1-\frac{1}{e}\right)d\sigma_l(e|h)< 0\]
    The profitable deviation is constructed by considering a menu that provides the same experiment to the high type and at the same cost. But the menu option for low type is changed and priced slightly higher. The low types's new experiment ($\widetilde{\sigma}_l$) has the same distribution for $e\in E\setminus\{e'\}$. But the probability of signal $e'$ in the low type's menu option is increased by shifting mass from the set $E^C$, until either $\frac{d\widetilde{\sigma}_h(e'|h)}{d\widetilde{\sigma}_l(e'|h)}= \frac{1}{e'l(\mu)}$ or $\int_E\left( 1-\frac{1}{e}\right)d\widetilde{\sigma}_l(e|h) = 0$. \\
    To construct such a deviation, note that $\int_Ed\sigma_l(e|h)<\int_E\frac{1}{e}d\sigma_l(e|h)< l(\mu)$, thus we have that $0< \int_{E^c}\frac{1}{e} d\sigma_l(e|h) < \int_{E^c} d\sigma_l(e|h)$. \\
    Whenever $\int_{E^c} d\sigma_l(e|l) - \frac{1}{e'}\int_{E^c}d\sigma_l(e|h) \geq 0$ then $\widetilde{\sigma}_l$ can be constructed such that for some $\kappa >0$; $\widetilde{\sigma}_l(e'|h) = \sigma_l(e'|h) + \kappa\int_{E^c}d\sigma_l(e|h)$, $\widetilde{\sigma}_l(0|l) = \sigma(0|l) + \kappa \left[ \int_{E^c} d\sigma_l(e|l) - \frac{1}{e'}\int_{E^c}d\sigma_l(e|h)\right] $ and $\widetilde{\sigma}_l(e|h) = \sigma_l(e|h)$ for all $e\not \in \{e',0\}$. 
    
    If $\int_{E^c} d\sigma_l(e|l) - \frac{1}{e'}\int_{E^c}d\sigma_l(e|h) < 0$ there exists some $E'\subset E^c$ such that $\int_{E'}d\sigma_l(e|h) > 0$ and $e\in E'$ implies $e> e'$, in this case for choose small enough $\kappa_1,\kappa_2> 0$ such that $\kappa_1\int_{E'} d\sigma_l(e|l) + \kappa_2 \int_{E_l} d\sigma_l(e|l) =\frac{1}{e'}\left( \kappa_1\int_{E'} d\sigma_l(e|h) + \kappa_2 \int_{E_l} d\sigma_l(e|h)\right)$. Define $\widetilde{\sigma}_l(e'|h) = \sigma_l(e'|h) + \kappa_1\int_{E'} d\sigma_l(e|h) + \kappa_2 \int_{E_l} d\sigma_l(e|h),~$  $\widetilde{\sigma}_l(e|h) = (1-\kappa_1)\sigma_l(e|h)$ for all $e \in E'$,  $\widetilde{\sigma}_l(e|h) = (1-\kappa_2)\sigma'_l(e|h)$ for all $e \in E_l$ and $\widetilde{\sigma}_l(e|h) = \sigma_l(e|h)$ for all $e\not \in \{e'\}\cup E'\cup E_l$. Thus in the optimal menu $\int_E\left( 1-\frac{1}{e}\right)d\sigma_l(e|h)\geq 0$.\end{proof}

\subsection{Proof of Corollary 1}

\textbf{Corollary 1:}

  If $m\in \mathcal{M}^r_E$ and $\text{ rev}(m)>0$ for some $E\subset [0,\infty]$ then $\int_E d\sigma^m_h(e|h)=1$.

\begin{proof}
    Following proposition 2, if $m\in \mathcal{M}^r_E$ then it solves the optimal program mentioned before. In particular, if $\int_Ed\sigma^m_h(e|h)<1$ then implementablity requires $\int_E\frac{1}{e}d\sigma^m_h(e|h) \leq \int_Ed\sigma^m_h(e|h) \leq 1-\epsilon $ for some $\epsilon>0$. Thus there exist $E_h,E_l\subset E^C$ such that $0<\int_{E_h}d\sigma_h^m(e|h),\int_{E_l}d\sigma_h^m(e|l)\leq\epsilon$. By lemma 5 we get that there exists $e'>1\in E$ (if $\infty\in E$ then let $e'=\infty$). If $\int_{E_h\cap E_l}\sigma_h^m(e|h), \int_{E_h\cap E_l}\sigma_h^m(e|l) > 0$ then the certifier has a profitable deviation thus $m\not\in \mathcal{M}^r_E$. To construct a profitable deviation of the certifier, we consider a menu $m'$ such that the low type's payment and experiment are equal to the low type's payment and experiment under the menu $m$. For the high type if $\frac{1}{e'}\int_{E_h\cap E_l}\sigma_h^m(e|h)\leq \int_{E_h\cap E_l}\sigma_h^m(e|l)$ let $\sigma^{m'}_h(e|h):=\sigma^m_h(e|h)$ whenever $e\in E\setminus\{e'\}$, $\sigma^{m'}_h(e'|h):= \sigma^{m}_h(e'|h) + \int_{E_h\cap E_l}\sigma_h^m(e|h)$ and $\rho^{m'}_h := \rho^{m}_h +\int_{E_h\cap E_l}\sigma_h^m(e|h)$. When $\frac{1}{e'}\int_{E_h\cap E_l}\sigma_h^m(e|h)> \int_{E_h\cap E_l}\sigma_h^m(e|l)$, let $\sigma^{m'}_h(e|h):=\sigma^m_h(e|h)$ whenever $e\in E\setminus\{e',\infty\}$,  $\sigma^{m'}_h(e'|h):= \sigma^{m}_h(e'|h) + e'\int_{E_h\cap E_l}\sigma_h^m(e|l)$, $\sigma^{m'}_h(\infty|h):= \sigma^{m}_h(\infty|h) + \left[ \int_{E_h\cap E_l}\sigma_h^m(e|h) - e'\int_{E_h\cap E_l}\sigma_h^m(e|l) \right]$, and $\rho_h^{m'}:= \rho^{m}_h +e'\int_{E_h\cap E_l}\sigma_h^{m}(e|h)$. If instead $\int_{E_h\cap E_l}\sigma_h^m(e|h)= 0$, then $\sigma_h^m(\infty|h), \sigma_h^m(0|l)>0$. Choose $\delta>0$ small enough such that $\sigma_h^m(\infty|h) > \delta,~ \sigma_h^m(0|l)>\frac{\delta}{e'}$. The certifier has a profitable deviation $m'$ such that $\sigma_l^{m'}:=\sigma_l^m$, $~\rho_l^{m'}:=\rho_l^{m}$, $~\sigma^{m'}_h(e|h):=\sigma^m_h(e|h)$ whenever $e\in E\setminus\{e'\}$, $\sigma^{m'}_h(e'|h):= \sigma_h^m(e'|h) + \delta$ and $\rho_h^{m'}= \rho_h^m +\delta$. The menu $m'\in \mathcal{M}_E$ for all the cases above by construction. Thus the corollary follows from contradiction.
\end{proof}
    

\subsection{Proof of Theorem 1}

\textbf{Theorem 1:}

If an equilibrium $((E_m)_{m\in\mathcal{M}}),m^*)\in \mathcal{E}_r$ is separating then $e^*:= \inf(E_{m^*}) \geq \frac{1}{\mu}$. Moreover, if for an equilibrium $((E_m)_{m\in\mathcal{M}}),m^*)\in \mathcal{E}_r$ it holds that $e^*:= \inf(E_{m^*}) > \frac{1}{\mu}$, then the equilibrium is separating.

\begin{proof}
    Sufficiency: Let $((E_m)_{m\in\mathcal{M}}),m^*)\in \mathcal{E}_r$. If $E_{m^*}$ is such that $e\in E_{m^*}$ implies that $e\geq 1$ then by applying proposition 2 we get that $m^*$ needs to be a solution to 
     \[ \max_{(\sigma_h^m(.|h),\sigma_l^m(.|h))\in \Delta([0,\infty])\times \Delta([0,\infty])}~~\mu\int_{E_{m^*}} d\sigma_h^m(e|h) + \int_{E_{m^*}}\left( \frac{1}{e}-\mu\right) d\sigma_l^m(e|h)\]
   subject to
   \[ \int_{E_{m^*}} \left(1-\frac{1}{e}\right)d\sigma_h^m(e|h) \geq \int_{E_{m^*}} \left(1-\frac{1}{e}\right)d\sigma_l^m(e|h)\geq 0\]
   \[d\sigma_h^m(e|h)\geq \frac{1}{el(\mu)}{d\sigma_l^m(e|h)}~~\text{for all } e\in E_{m^*}\]

   The non-negativity constraint in the IC conditions is irrelevant as $e>1$ for all $e\in E_{m^*}$. Now notice that if $e^*\geq \frac{1}{\mu}$ then the revenue of the certifier is decreasing in $d\sigma_l^m(e|h)$ for all $e\in E_{m^*}$. This means that the low type's menu option assigns zero probability to the set $E_{m^*}$. In particular, the revenue maximizing menu $m^*$ is such that $\int_{E_{m^*}}d\sigma^{m^*}_h(e|h) = 1$, $\rho_h^{m^*}=1$, $\sigma^{m^*}_l = \Phi$, $\rho_l^{m^*}=0$. This menu is separating. Notice when $e^* > \frac{1}{\mu}$ then the revenue is strictly decreasing in $\sigma^{m^*}_l(e|h)$ for all $e\in E_{m^*}$ thus the menu with $\int_{E_{m^*}}d\sigma^{m^*}_h(e|h) = 1$, $\rho_h^{m^*}=1$, $\sigma^{m^*}_l = \Phi$, $\rho_l^{m^*}=0$ is uniquely optimal. 
   
   For necessity, let $((E_m)_{m\in\mathcal{M}}),m^*)\in \mathcal{E}_r$ be a separating equilibrium. I will show that if there exists $e\in E_{m^*}$ such that $e<\frac{1}{\mu}$ then $m^*\not\in \mathcal{M}_{E_{m^*}}^r$. \\
   Case I: There exists $0<e_l<1\in E_{m^*}$. As $m^*$ is separating so by lemma 5 we get that there exists some $e_h>1\in E_{m^*}$. Now consider the following menu: $ \chi = ((\sigma_\theta,\rho_\theta))_{\theta\in \{l,h\}}$. Where $\sigma^{\chi}_h(e_h|h)=\frac{e_h(1-e_l)}{e_h-e_l}$, $\sigma^{\chi}_h(e_l|h) = \frac{e_l(e_h-1)}{e_h-e_l}$, $\rho_h = 1$ and  $\sigma^{\chi}_l(e_h|h)=l(\mu)e_l\frac{e_h(1-e_l)}{e_h-e_l}$, $\sigma^{\chi}_l(e_l|h) = l(\mu)e_l\frac{e_l(e_h-1)}{e_h-e_l}$, $\rho_l = e_ll(\mu)$. Observe that $\chi\in \mathcal{M}_{E_{m^*}}$, and $\text{rev}(\chi) = \mu + (1-\mu)l(\mu)e_l$. The claim follows from noting that the revenues from a separating menu is bounded above by $\mu$.\\
   Case II: $e\in E$ implies $e>1$ and there exists $e_h\in E$ such that $\frac{1}{\mu}> e_h>1$. If $e_h\geq \frac{1}{l(\mu)}$, the certifier can improve revenue by offering the same menu option to both types. This is $\sigma_h=\sigma_l$ where $\sigma_h(e_h|h)=1$, $\sigma_h(e_h|l)=\frac{1}{e_h}$ and $\rho_h=\rho_l = \frac{1}{e_h}$. \\
   If $e_h<\frac{1}{l(\mu)}$. The certifier can improve the revenue by offering a menu such that $\sigma_h(e_h|h)=1$, $\sigma_l(e_h|h)=e_hl(\mu)$ and $\rho_h = 1-l(\mu)(e_h-1)$, $\rho_l=l(\mu)$. 
\end{proof}

\subsection{Proof of Proposition 1}

\textbf{Proposition 1:}

    The set $\mathcal{E}_n\subset \mathcal{E}_r$. If $((E_m)_{m\in\mathcal{M}},m^*)\in \mathcal{E}_n$ then $\sigma_h^{m^*}=\sigma_l^{m^*}$ and $\sigma_h^{m^*}\left(\frac{1}{l(\mu)}|h\right)=1$. Moreover, $((E_m)_{m\in\mathcal{M}},m^*)\in \mathcal{E}_n$ is separating (if and) only if $\mu (<)\leq 2-\frac{1}{\pi^*}$.

\begin{proof}
    Any equilibrium in $\mathcal{E}_n$ satisfies the refinement and thus is a subset of $\mathcal{E}_r$. When $((E_m)_{m\in\mathcal{M}},m^*)\in \mathcal{E}_n$, then proposition 2 and observation 1 show that the equilibrium payments and the equilibrium experiments solves 
    \[ \max_{(\sigma_h^m(.|h),\sigma_l^m(.|h))\in \Delta([0,\infty])\times \Delta([0,\infty])} ~~\mu\int_{\frac{1}{l(\mu)}}^\infty d\sigma_h^m(e|h) + \int_{\frac{1}{l(\mu)}}^\infty\left( \frac{1}{e}-\mu\right) d\sigma_l^m(e|h)\]
   subject to
   \[ \int_{\frac{1}{l(\mu)}}^\infty \left(1-\frac{1}{e}\right)d\sigma_h^m(e|h) \geq \int_{\frac{1}{l(\mu)}}^\infty \left(1-\frac{1}{e}\right)d\sigma_l^m(e|h)\geq 0\]
   \[d\sigma_h^m(e|h)\geq \frac{1}{el(\mu)}{d\sigma_l^m(e|h)}~~\text{for all } e\in E\]
   The problem simplifies by noting that $ \int_{\frac{1}{l(\mu)}}^\infty \left(1-\frac{1}{e}\right)d\sigma_h^m(e|h),~\int_{\frac{1}{l(\mu)}}^\infty \left(1-\frac{1}{e}\right)d\sigma_l^m(e|h)\geq 0$ always holds. 
   Note that $\mu\geq l(\mu)$ if and only if $2-\frac{1}{\pi^*}\geq \mu$. 

    If $\mu < 2-\frac{1}{\pi^*}$ then by theorem 1 we get that the equilibrium menu $m^*$ is given by:
    \[ \int_{\frac{1}{l(\mu)}}^\infty d\sigma_h^{m^*}(e|h)=1,~\rho_h=1, \text{ and } \sigma^{m^*}_l=\Phi, \rho_l=0\]
    If $\mu>2-\frac{1}{\pi^*}$ then the revenue is strictly increasing in probability $\int_{\frac{1}{l(\mu)}}^{\frac{1}{\mu}}d\sigma^m_l(e|h)$, the equilibrium menu is derived by noting that $\frac{1}{e}-\mu$ is decreasing in $e$ and that $l(\mu)e >1$, $\int_{l(\mu)}^1 d\sigma^{m^*}_l(e|h) \leq \int_{l(\mu)}^1 d\sigma^{m^*}_h(e|h) =1$.
    \[\sigma_h^{m^*}=\sigma_l^{m^*};~\sigma_h^{m^*}\left(\frac{1}{l(\mu)}|h\right)=1 \text{ and } \rho_h=\rho_l=l(\mu)\]
    If $\mu = 2-\frac{1}{\pi^*}$, then either of the above menus can be offered in equilibrium. 
\end{proof}

\subsection{Proof of Theorem 2}

\textbf{Theorem 2:}

    The equilibrium $((E_m)_{m\in\mathcal{M}},m^*)\in \mathcal{E}_r$ yields the highest ex-ante expected payoff for the sender among all equilibrium in $\mathcal{E}_r$ only if $\min(E_{m^*})=\frac{1}{\mu}$. Moreover, the equilibrium menu $m^*$ is such that $\sigma_h^{m^*}\left(\frac{1}{\mu}~|~h\right) = 1.$

\begin{proof}
    By proposition 2 we know the expected surplus of the sender is the information rent earned by the high type. Let $e^* = \inf(E)$. When $e>1$ for all $e\in E$ then for any $((E)_{m\in\mathcal{M}},m^*)\in \mathcal{E}_r$ yields the following surplus to the high type sender: $0$ if $e^*> \frac{1}{\mu}$ by theorem 1, either $0$ or $1-\mu$ if $e^* = \frac{1}{\mu}$, $1-\frac{1}{e^*}$ if $\frac{1}{l(\mu)} < e^* < \frac{1}{\mu}$ and $(e^*-1)l(\mu)$ if $e^* \leq\min\left(\frac{1}{l(\mu)} \frac{1}{\mu}\right)$. Thus $\min(E)=\frac{1}{\mu}$ achieves the highest surplus for the high type. In particular, the maximum possible surplus is $1-\mu$ if $\frac{1}{l(\mu)}<\frac{1}{\mu} $  and the surplus is $\frac{l(\mu)}{\mu}(1-\mu)$ if $\frac{1}{l(\mu)} \geq \frac{1}{\mu} $.
    
Now we look at equilibrium $((E'_{m})_{m\in\mathcal{M}}, m')\in \mathcal{E}_r$ such that there be some $e_l<1 \in  E'_{m'}$. By corollary 1 we know that $\int_{E'_{m'}}d\sigma_h^m(e|h) =1$, then by lemma 5 there exists some $e_h>1 \in E$.\\ 
First, consider the case when $\frac{1}{l(\mu)}\geq \frac{1}{\mu}$. By construction similar to theorem 1 case 1 we get that the certifier can guarantee himself at least the revenue
\[\mu\left[ 1 - \int_{E'_{m'}}\left(1-\frac{1}{e}\right)d\sigma^{m'}_l(e|h)\right] +(1-\mu)\int_{E'_{m'}}d\sigma^{m'}_l(e|l) \geq \mu + e_ll(\mu)(1-\mu)\]
    \[\implies (1-\mu)\int_{E'_{m'}}d\sigma^{m'}_l(e|l) - e_ll(\mu)(1-\mu) \geq \mu\int_{E'_{m'}}\left( 1-\frac{1}{e}\right) d\sigma^{m'}_l(e|h)\]
    Where the RHS is the expected rent earned by the sender in the revenue-maximizing menu wrt ${E'_{m'}}$. Fix an equilibrium $((E_m)_{m\in\mathcal{M}},m^*) \in \mathcal{E}_r$ such that $\min(E_{m^*}) = \frac{1}{\mu}$, by first part of the proof the surplus earned by the high type is $(1-\mu)\frac{l(\mu)}{\mu}$. The result follows by noting that 
    \[(1-\mu)\int_{E'_{m'}}d\sigma^{m^*}
    _l(e|l) - e_ll(\mu)(1-\mu) \leq (1-\mu)l(\mu)(1-e_l)<(1-\mu)l(\mu)\] 

   Now, consider the case when $\frac{1}{l(\mu)} < \frac{1}{\mu}$. Fix an equilibrium $((E_m)_{m\in\mathcal{M}},m^*) \in \mathcal{E}_r$ such that $\min(E_{m^*}) = \frac{1}{\mu}$, from discussion in the first part of the proof we know that $\int_{E_{m^*}}d\sigma_h^{m^*}(e|h) = 1$, $\int_{E^{m^*}}d\sigma_l^{m^*}(e|h) = 1$, and $\int_{E^{m^*}}d\sigma_l^{m^*}(e|l) = \mu$.
   Consider for contradiction that there exists an equilibrium $((E'_{m})_{m\in\mathcal{M}},m')\in \mathcal{E}_r$ such that $e_l<1\in E'_{m'}$ and 
   \[\int_{E'_{m'}}\left(1-\frac{1}{e}\right)d\sigma_l^{m'}(e|h) > \int_{E_{m^*}}\left(1-\frac{1}{e}\right)d\sigma_l^{m^*}(e|h) >0\]
   As $\int_{E^{m^*}}d\sigma_l^{m^*}(e|h) = 1$, this implies that 
   \[\int_{E'_{m'}}\frac{1}{e}d\sigma_l^{m'}(e|h) < \int_{E_{m^*}}\frac{1}{e}d\sigma_l^{m^*}(e|h) \]
   By construction similar to theorem 1 case 1 we get that the certifier can guarantee himself at least the revenue
   \[\mu\left[ 1 - \int_{E'_{m'}}\left(1-\frac{1}{e}\right)d\sigma^{m'}_l(e|h)\right] +(1-\mu)\int_{E'_{m'}}d\sigma^{m'}_l(e|l) \geq \mu + e_ll(\mu)(1-\mu) > \int_{E^{m^*}}d\sigma_l^{m^*}(e|l)\]
    Rearranging the terms we get that 
    \[-\mu\int_{E'_{m'}}\left(1-\frac{1}{e}\right)d\sigma_l^{m'}(e|h) + (1-\mu) \int_{E'_{m'}}\frac{1}{e}d\sigma_l^{m'}(e|h) \]
    \[> -\mu \int_{E_{m^*}}\left(1-\frac{1}{e}\right)d\sigma_l^{m^*}(e|h)+ (1-\mu)\int_{E_{m^*}}\frac{1}{e}d\sigma_l^{m^*}(e|h)\]

    \[\implies \mu\left[ \int_{E_{m^*}}\left(1-\frac{1}{e}\right)d\sigma_l^{m^*}(e|h) - \int_{E'_{m'}}\left(1-\frac{1}{e}\right)d\sigma_l^{m'}(e|h)\right] \]
    \[\geq (1-\mu) \left[\int_{E_{m^*}}\frac{1}{e}d\sigma_l^{m^*}(e|h) -  \int_{E'_{m'}}\frac{1}{e}d\sigma_l^{m^*}(e|h)\right] \]

    This is a contradiction as RHS is positive and the LHS is negative. 
   \end{proof}

\subsection{Proof of Theorem 3}

\textbf{Theorem 3:}
      If $((E_m)_{m\in\mathcal{M}}, m^*)\in \mathcal{E}_r$ such that $E_{m^*}$ is countable, then there exists an outcome equivalent equilibrium  $((E'_m)_{m\in\mathcal{M}}, m')\in \mathcal{E}_r$ such that $m'\in \textbf{cvx}(\mathcal{T}(E_{m^*}))$.  \\
      
Combining corollary 1, the constraint $\int_E\left(1-\frac{1}{e}\right) d\sigma_{\theta}(e|h) \geq 0$ and remark 2. We get that it suffices to consider menus such that $\supp(\sigma_{\theta}(.|l))\cap E^C \subset \{0,1\}$. With this in mind, I restrict the space of choices from $\Delta([0,\infty])\times \Delta([0,\infty])$ to $\Lambda$, where
\[\Lambda : = \{(\eta_h,\eta_l)  \in \Delta([0,\infty])\times \Delta([0,\infty]) | \supp(\eta_{\theta})\cap E^C =\{1\}\}\]
 $\Lambda$ is a closed and convex set\footnote{The topology here is product topology, where the topology on each component is the topology of weak convergence.}. 



For each $E\subset [0,\infty]$. The set of \textbf{feasible}\footnote{The requirement of feasibility wrt $E$ is stronger than obedience wrt $E$ as feasibility builds in revenue-maximizing properties that are discussed in proposition 2.} signal distributions wrt $E$ is given by $\mathcal{K}_E\subset \Lambda$. Where $~\mathcal{K}_E$ is the set of signal distributions $(\eta_h,\eta_l) \in \Lambda$ such that  
    \[\int_E \left(1-\frac{1}{e}\right)d\eta_h(e) \geq \int_E \left(1-\frac{1}{e}\right)d\eta_l(e)\]
    \[\int_E \left(1-\frac{1}{e}\right)d\eta_l(e)\geq 0\]
    \[\int_Ed\eta_h(e)=1\]
    and 
   \[\int_{E'}d\eta_h(e)\geq \int_{E'}\frac{1}{el(\mu)}{d\eta_l(e)}~~\text{for all } E'\subset E \] 

The first claim establishes the existence, in terms of restrictions on $E$, of a maximizer of some linear functional over $\mathcal{K}_E$.

\begin{claim}
    For any $
    \epsilon >0$ and $E\subset [\epsilon,\infty]\setminus \{1\}$ closed, the set $\mathcal{K}_E$ is compact and convex.
\end{claim}
\begin{proof}
    Follows from standard arguments.
\end{proof}

A consequence of the claim and Krein–Milman theorem is that the set of extreme points $ex(\mathcal{K}_E)$ is nonempty. As a convention if some functional $f$ doesn't attain its maximum in the set $ \mathcal{K}_E $, then $\argmax_{\mathcal{K}_E}\{f(\eta_h,\eta_l)\} = \varnothing$.
For all the following statements in the proof, I assume that $E$ is countable. In particular, the support of all the relevant distributions is discreet\footnote{The proofs can be extended to continuous distribution, the arguments presented rely on discreet support distributions mostly for exposition reasons.}.  

Let $\mathcal{T}(\mathcal{K}_E) := \{(\eta_h,\eta_l)\in \mathcal{K}_E| ~|\supp(\eta_h)\cap E| \leq 3,~~|\supp(\eta_l)\cap E| \leq 2\}$. 

\begin{claim}
    If $E\cap [0,1]=\varnothing$ then $\argmax_{\mathcal{K}_E}\{\mu\int_Ed\eta_h(e)+ \int_{E}\left( \frac{1}{e}-\mu\right) d\eta_l(e)\}\subset\mathcal{T}(\mathcal{K}_E)$.
\end{claim}

\begin{proof}
    Let $(\eta_h^*,\eta^*_l)\in \argmax_{\mathcal{K}_E}\{\mu\int_Ed\eta_h(e)+ \int_{E}\left( \frac{1}{e}-\mu\right) d\eta_l(e)\}$. By solving the optimization problem in proposition 2 (see proof of theorem 1 for example), we get that $|\supp(\eta^*_h)\cap E| >1$ only if $\eta^*_l(1) =1$, in which case we can write  $(\eta^*_h,\eta^*_l) = \sum_{e\in E}\eta_h(e)(\eta^e_h,\eta^e_l)$. Where $\eta^e_l=\eta^*_l$ and $\eta^e_h(e) = 1$ for all $e\in E$. 
\end{proof}

In particular if $(\eta'_h,\eta'_l)\in \argmax_{\mathcal{K}_E}\{\mu\int_Ed\eta_h(e)+ \int_{E}\left( \frac{1}{e}-\mu\right) d\eta_l(e)\}$ are such that $\supp(\eta_h)\cap [0,1]\cap E=\varnothing$ then $(\eta'_h,\eta'_l)\in \argmax_{\mathcal{K}_{E\setminus[0,1]}}\{\mu\int_Ed\eta_h(e)+ \int_{E}\left( \frac{1}{e}-\mu\right) d\eta_l(e)\}$.

\begin{claim}
For any $(\eta_h,\eta_l) \in \argmax_{K_E}\{\mu\int_Ed\widetilde{\eta}_h(e)+ \int_{E}\left( \frac{1}{e}-\mu\right) d\widetilde{\eta}_l(e)\}$. If $\supp(\eta_h) \cap E\cap [0,1]\neq \varnothing$ then $\supp(\eta_l) \cap E\cap [0,1] \neq \varnothing$.
\end{claim}

\begin{proof}
    Consider for contradiction $\supp(\eta_h) \cap E\cap [0,1]\neq \varnothing$ and $\supp(\eta_l) \cap E\cap [0,1] = \varnothing$. Let $E_l\subset E\cap [0,1]$ be such that $\int_{E_l}d\eta_h(e)>0$, fix some $e_l\in E_l$.
    By assumption $\int_{E_l}d\eta_l(e)=0$.  Let $E_h\subset E\cap [1,\infty]$ be such that $\int_{E_h}d\eta_l(e)>0$, fix some $e_h\in E_h$.
    
 First, consider the case when $\eta_l(1) = 0$, then we get that $\int_Ed\eta_h(e) = \int_E d\eta_l(e) =1$ and $\int_E\frac{1}{e}d\eta_l \geq \int_E\frac{1}{e}d\eta_h$. As $\frac{d\eta_l(e)}{d\eta_h(e)}\leq el(\mu)$ for all $e\in E$, we get that $\int_E\frac{1}{e}d\eta_l(e)\leq l(\mu)$. Now construct $\eta_h',\eta_l'$ such that $\eta'_h= \eta_h$ and $\eta'_l(e)  = \eta_l(e)$ for all $e\not \in \{e_l,~e_h\}$. Let $\eta'_l(e_h) = \eta_l(e_h) - \epsilon$, 
 $\eta'_l(e_l) = \eta_l(e_l) + \epsilon$,
 and $1-\int_E\frac{1}{e}d\eta'_l(e) = 1-\int_E\frac{1}{e}d\eta_l(e) - \epsilon \left(\frac{1}{e_l}-\frac{1}{e_h}\right)$.  
 For small enough $\epsilon>0$ we get that $(\eta_h',\eta_l')\in \mathcal{K}_E$ and 
    \[ \mu\int_Ed\eta'_h(e)+ \int_{E}\left( \frac{1}{e}-\mu\right) d\eta'_l(e) - \mu\int_Ed\eta_h(e)- \int_{E}\left( \frac{1}{e}-\mu\right) d\eta_l(e)\]
    \[ = \mu \epsilon  \left(\frac{1}{e_l} - \frac{1}{e_h}\right) > 0\]
    This is a contradiction, thus it must be that $\int_{E_l}d\eta_l(e) > 0$. 
 
 Now, consider the case when $\eta_l(1) > 0 $. Construct $\eta_h',\eta_l'$ such that for all $e\not \in \{e_h, 1\}\cup E_l$ we have $\eta_{\theta}'(e) = \eta_{\theta}(e)$. Let
    $\eta_h'(e_h) = \eta_h(e_h) + \epsilon\int_{E_l}d\eta_h(e)$,
    $\eta_l'(e_h) = \eta_l(e_h) + \epsilon e_hl(\mu)\int_{E_l}d\eta_h(e)$,
    $\eta_h'(e) = (1-\epsilon)\eta_h(e)$ for all $e\in E_l$,
    $\eta_l'(e) = \eta_l(e)$ for all $e\in E_l\setminus\{e_l\}$, 
    $\eta'_l(e_l) = \eta_l(e_l) +\epsilon l(\mu) \frac{(e_h-1)e_l}{1-e_l}\int_{E_l}d\eta_h(e)$, 
    $\eta'_h(1)=\eta_h(1)$,
    $1-\int_{E\cup \{1\}}\frac{1}{e}d\eta_h'(e) = 1-\int_{E\cup \{1\}}\frac{1}{e}d\eta_h(e) + \epsilon\int_{E_l}\left(\frac{1}{e}-\frac{1}{e_h}\right)d\eta_h(e)$,
    $1-\int_{E\cup \{1\}}\frac{1}{e}d\eta_l'(e) = 1-\int_{E\cup \{1\}}\frac{1}{e}d\eta_l(e)$,
    and $\eta'_l(1)=\eta_l(1) - \epsilon l(\mu)\frac{e_h-e_l}{1-e_l}\int_{E_l}d\eta_h(e)$. 
   For small enough $\epsilon>0$ we get that $(\eta_h',\eta_l')\in \mathcal{K}_E$ and 
    \[ \mu\int_Ed\eta'_h(e)+ \int_{E}\left( \frac{1}{e}-\mu\right) d\eta'_l(e) - \mu\int_Ed\eta_h(e)- \int_{E}\left( \frac{1}{e}-\mu\right) d\eta_l(e)\]
    \[ = (1-\mu)\epsilon l(\mu)\frac{e_h-e_l}{1-e_l}\int_{E_l}d\eta_h(e) > 0\]
    This is a contradiction, thus it must be that $\int_{E_l}d\eta_l(e) > 0$. 
\end{proof}

 For the rest of the claims let $\supp(\eta_l)\cap E\cap [0,1]\neq\varnothing$.
 
 Partition $\mathcal{K}_E$ into $K^E_0$ and $K^E_1$. Where $K^E_0:=\{(\eta_h,\eta_l)\in \mathcal{K}_E| \int_E\left(1-\frac{1}{e}\right) d\eta_l(e) =0\}$, and $K_1^E=  \mathcal{K}_E\setminus K^E_1$. 

 \begin{claim}
    For any $(\eta_h,\eta_l)\in  \argmax_{\mathcal{K}_E}\{\mu\int_Ed\widetilde{\eta}_h(e)+ \int_{E}\left( \frac{1}{e}-\mu\right) d\widetilde{\eta}_l(e)\}$. If $\supp(\eta_l)\cap E\cap [0,1]\neq\varnothing$ and $\int_E\left(1-\frac{1}{e}\right)d\eta_l(e) > 0 $, then $\supp(\eta_l)\cap E\cap (\frac{1}{l(\mu)},\infty] = \varnothing$ or $|\supp(\eta_l)\cap E\cap [1,\infty]| = 1$.
\end{claim}

\begin{proof}
    Note that $\supp(\eta_l)\cap E\cap [0,1]\neq\varnothing$ implies that there is some $e_l\in E\cap [0,1]$ such that $\eta_h(e_l)>\eta_l(e_l)>0$.  By assumption we have $\int_E\left(1-\frac{1}{e}\right) d\eta_l(e) >0$. As $\int_E\left(1-\frac{1}{e}\right) (d\eta_h(e)-d\eta_l(e))\geq 0$, we get that  $\int_E\left(1-\frac{1}{e}\right) d\eta_h(e) >0$. For contradiction let there be some $e',e_h\in E\cap[1,\infty]$ such that $e'<e_h$, $\eta_l(e_h) > 0$, and  $\frac{1}{l(\mu)}<e_h$. 
    
    Now construct $\eta'_h,\eta'_l$ such that $\eta'_{\theta}(e) = \eta_{\theta}(e)$ for $e\not \in \{e_h,e',0,1\}$.
    Let $\eta_h'(e_h)=\eta_h(e_h) - \epsilon$, 
$\eta_h'(e')=\eta_h(e')+\epsilon$,
$1-\int_Ed\eta'_h(e) = 1-\int_Ed\eta_h(e) - \left(\frac{1}{e'}-\frac{1}{e_h}\right)\epsilon$, 
    $\eta_l'(e_h)=\eta_l(e_h) - e_hl(\mu)\epsilon$, 
$\eta_l'(e')=\eta_l(e')+e'l(\mu)\epsilon$, 
$\eta_l'(1)=\eta_l(1)+\epsilon l(\mu)(e_h -  e')$, and
$1-\int_Ed\eta'_l(e) = 1-\int_Ed\eta_l(e) - l(\mu)\epsilon(e_h-e')$.
    For small enough $\epsilon> 0$, we get $(\eta'_h,\eta'_l)\in \mathcal{K}_E$ as
    \[\left(1-\frac{1}{e_h}\right)\epsilon(e_hl(\mu)-1) -\left(1-\frac{1}{e'}\right)\epsilon (l(\mu)e'-1)\]
    \[ = \epsilon(e_h-e')\left(l(\mu) -\frac{1}{e_he'}\right) > 0\]
   Finally, note that 

   \[\mu\int_Ed\eta'_h(e)+ \int_{E}\left( \frac{1}{e}-\mu\right) d\eta'_l(e) - \mu\int_Ed\eta^*_h(e)+ \int_{E}\left( \frac{1}{e}-\mu\right) d\eta^*_l(e)\]
    
    \[= \mu\epsilon l(\mu)(e_h-e') >0\] \end{proof}

A direct consequence of claim 4 is that for any $(\eta_h,\eta_l) \in \argmax_{K_E}\{\mu\int_Ed\widetilde{\eta}_h(e)+ \int_{E}\left( \frac{1}{e}-\mu\right) d\widetilde{\eta}_l(e)\}$, if $\supp(\eta_l)\cap E\cap [0,1]\neq\varnothing$, then $\eta_l(1) > 0$. 

 \begin{claim}
     For any $(\eta_h,\eta_l) \in \argmax_{K_E}\{\mu\int_Ed\widetilde{\eta}_h(e)+ \int_{E}\left( \frac{1}{e}-\mu\right) d\widetilde{\eta}_l(e)\}$. Such that $\supp(\eta_l)\cap E\cap [0,1]\neq\varnothing$. If $|\supp(\eta_h)\cap E \cap [1,\infty]|\geq 2$ then for any $e',e_h \in \supp(\eta_h)\cap E \cap [1,\infty]$, $e'<e_h$ implies that $\frac{\eta_l(e_h)}{\eta_h(e_h)} = 0~or~e_hl(\mu)$, and if $\frac{\eta_l(e_h)}{\eta_h(e_l)} = e_hl(\mu)$ then $\frac{\eta_l(e')}{\eta_h(e')} = e'l(\mu)$. 
 \end{claim}

 \begin{proof}
     Note that $\supp(\eta_l)\cap E\cap [0,1]\neq\varnothing$ implies that there is some $e_l\in E\cap [0,1]$ such that $\eta_h(e_l)>\eta_l(e_l)>0$. In particular, by claim 4 we have $\eta_l(1) > 0$.
     
     First, for contradiction assume that $0<\frac{\eta_l(e_h)}{\eta_h(e_h)}< e_h l(\mu)$. Construct $(\eta_h',\eta_l')$ such that for all $e\not \in \{1,0,e_h,e_l,e'\}$ we have $\eta_{\theta}'(e) = \eta_\theta(e)$.
     Let $\eta_h'(e_h) = \eta_h(e_h) - \epsilon$, 
     $\eta_h'(e_l) = \eta_h(e_l) - \delta$, $\eta_h'(e') = \eta_h(e') + \epsilon + \delta$,
     $\eta_l'(e_h) = \eta_l(e_h) - \epsilon\psi$, 
     $\eta_l'(e_l) = \eta_l(e_l) - e_ll(\mu) \delta$, 
     $\eta_l'(e') = \eta_l(e') + \epsilon\psi + e_ll(\mu)\delta + \kappa$,
     and $\eta_l'(1) = \eta_l(1) - \kappa$.
     Where $\kappa = \frac{1}{e'-1}(\epsilon\psi\left(1-\frac{e'}{e_h}\right) -\delta l(\mu) (e'-e_l))$,
     $\psi > \frac{\eta_l(e_h)}{\eta_h(e_h)}$,
     $\epsilon > \delta\frac{e'-e_l}{e_h-e'}\frac{l(\mu)e_h}{\psi}$ and $\delta > 0$.
     By choosing $\psi$, $\epsilon$, and $\delta$ close enough to the lower bounds mentioned respectively we get that $(\eta'_h,\eta_l')\in \mathcal{K}_E$, by construction we see that 
     \[ \mu\int_Ed\eta'_h(e)+ \int_{E}\left( \frac{1}{e}-\mu\right) d\eta'_l(e) - \mu\int_Ed\eta_h(e)- \int_{E}\left( \frac{1}{e}-\mu\right) d\eta_l(e)\]
     \[= (1-\mu)\kappa>0\]
     Now consider $e'<e_h$ where $\frac{\eta_l(e')}{\eta_h(e')}\neq e'l(\mu)$. Construct $\eta_h',\eta_l'$ such that $\eta_h'(e)=\eta_h(e)$ for all $e\in [0,\infty]$. For $e\not\in \{e_h,e',e_l,1\}$ let $\eta_l'(e) = \eta_l(e)$. 
     Let $\eta_l'(e_h) = \eta_l(e_h)-\epsilon$, 
     $\eta_l'(e_l) = \eta_l(e_l)-\delta$, $\eta_l'(1) = \eta_l(1)-\kappa$
     and $\eta_l'(e') = \eta_l(e')+\epsilon+\delta+\kappa$.
     Where $\kappa = \frac{e'}{e'-1}\left(\epsilon\left( \frac{1}{e'}-\frac{1}{e_h}\right) - \delta\left(\frac{1}{e_l}-\frac{1}{e'}\right)\right)$,
     $\epsilon> \delta\frac{(e'-e_l)e_h}{(e_h-e')e_l}$
     and $\delta>0$.
     By choosing $\epsilon$ and $\delta$ close enough to the lower bounds mentioned respectively we get that $(\eta'_h,\eta_l')\in \mathcal{K}_E$, by construction we see that 
     \[ \mu\int_Ed\eta'_h(e)+ \int_{E}\left( \frac{1}{e}-\mu\right) d\eta'_l(e) - \mu\int_Ed\eta_h(e)- \int_{E}\left( \frac{1}{e}-\mu\right) d\eta_l(e)\]
     \[= (1-\mu)\kappa>0\] \end{proof}

Following claim 5, I proceed by dividing the proof into two cases.
First, I consider $(\eta_h,\eta_l) \in \argmax_{K_E}\{\mu\int_Ed\widetilde{\eta}_h(e)+ \int_{E}\left( \frac{1}{e}-\mu\right) d\widetilde{\eta}_l(e)\}$, such that $\supp(\eta_l)\cap E\cap [0,1]\neq\varnothing$
and $ \frac{\eta_l(e)}{\eta_h(e)}=el(\mu)$ for all $e\in E\cap [1,\infty]$. 
Second, I consider the case $(\eta_h,\eta_l) \in \argmax_{K_E}\{\mu\int_Ed\widetilde{\eta}_h(e)+ \int_{E}\left( \frac{1}{e}-\mu\right) d\widetilde{\eta}_l(e)\}$, such that $\supp(\eta_l)\cap E\cap [0,1]\neq\varnothing$
and $ \frac{\eta_l(e)}{\eta_h(e)}<el(\mu)$ for some $e \in E\cap [1,\infty]$. Before proceeding I will first prove some claims that simplify the optimization problem.

 \begin{claim}
      For any $(\eta_h,\eta_l) \in \argmax_{K_E}\{\mu\int_Ed\widetilde{\eta}_h(e)+ \int_{E}\left( \frac{1}{e}-\mu\right) d\widetilde{\eta}_l(e)\}$, such that $\supp(\eta_l)\cap E\cap [0,1]\neq\varnothing$. If $\int_E\left(1-\frac{1}{e}\right) d\eta_l(e) = 0 $ then $\int_E\left(1-\frac{1}{e}\right) d\eta_l(e) = \int_E\left(1-\frac{1}{e}\right) d\eta_l(e) $.
 \end{claim}
\begin{proof}
    Fix some $(\eta_h,\eta_l) \in \argmax_{K_E}\{\mu\int_Ed\widetilde{\eta}_h(e)+ \int_{E}\left( \frac{1}{e}-\mu\right) d\widetilde{\eta}_l(e)\}$ such that $\supp(\eta_l)\cap E\cap [0,1]\neq\varnothing$. The proof follows in two steps: 
    
    The first argument is similar to the last part of the proof of proposition 2. 
    We want to show that if $\int_E\left(1-\frac{1}{e}\right) d\eta_l(e) = 0 $ then there exist some $e_h\in \supp(\eta_h)\cap E\cap [1,\infty]$ such that $\frac{\eta_l(e_h)}{\eta_h(e_h)} < e_hl(\mu)$.
Whenever $\int_E\left(1-\frac{1}{e}\right) d\eta_l(e) = 0 $ we have that there exists $E_l \subset \supp(\eta_l)\cap E\cap [0,1]$ such that $\int_{E_l} d\eta_l(e)>0$. 
Assume for contradiction that for all $e\in \supp(\eta_h)\cap E\cap [1,\infty]$ we have $\frac{\eta_l(e_h)}{\eta_h(e_h)} = el(\mu)$. 
Then it follows that
\[l(\mu)\int_{E\cap[1,\infty]}(e-1)d\eta_h(e) = \int_{E\cap[0,1]}\left(\frac{1}{e}-1\right) d\eta_l(e)  \leq l(\mu) \int_{E\cap [0,1]} (1-e) d\eta_h(e)\]
Feasibility further requires that 
\[ \int_{E\cap[0,1]} \frac{1}{e}d\eta_h(e) + \int_{E\cap[1,\infty]}\frac{1}{e}d\eta_h(e) \leq 1\]
\[\implies  \int_{E\cap[0,1]} \left(\frac{1}{e}-1\right)d\eta_h(e) \leq \int_{E\cap[1,\infty]}\left(1-\frac{1}{e}\right)d\eta_h(e)\]
Now note that $\int_{E\cap[0,1]}\left(\frac{1}{e}-1\right) d\eta_h(e) > \int_{E\cap[0,1]}(1-e)d\eta_h(e)$ and $ \int_{E\cap[1,\infty]}\left(1-\frac{1}{e}\right)d\eta_h(e) < \int_{E\cap[1,\infty]}(e-1)d\eta_h(e)$. Thus, we have a contradiction.

For the second argument, we want to show that if $0= \int_E\left(1-\frac{1}{e}\right)\eta_l< \int_E\left(1-\frac{1}{e}\right)\eta_h $ then for any $E_h\subset \supp(\eta_h)\cap E \cap [1,\infty]$, we have $\int_{E_h}d\eta_l(e) = l(\mu)\int_{E_h}ed\eta_h(e)$. 
Assume for contradiction that there is some $E_h \subset [1,\infty] \cap E$ such that $\int_{E_h}d\eta_h(e) > 0 $ and $\int_{E_h}d\eta_l(e) < l(\mu)\int_{E_h}ed\eta_h(e)$, in particular $E_h$ can be chosen such that $\frac{d\eta_l(e)}{d\eta_h(e)} < el(\mu)$ for all $e\in E_h$. Also, note that there is some $e_l\in E\cap [0,1]$. \\
Construct $\eta_h',\eta_l'$ such that for all
$e\not\in E_h \cup \{e_l,1,0\}$ we have $\eta_\theta'(e) = \eta_\theta(e)$. 
Define $\eta_h'(e) = (1-\epsilon)\eta_h(e)$ for all $e\in E_h$, 
$\eta_h'(e_l) = \eta_h(e_l) +\epsilon\int_{E_h} d\eta_h(e)$,
$1-\int_Ed\eta_h'(e) = 1-\int_Ed\eta_h(e) -\epsilon\int_{E_h}\left( \frac{1}{e_l}-\frac{1}{e}\right) d\eta_h(e)$,
$\eta_l'(e) =  \left(1+\epsilon l(\mu)\frac{e(1-e_l)}{e-1}\frac{\eta_h(e)}{\eta_l(e)}\right) \eta_l(e) $ for all $e\in E_h$, 
$\eta_l'(e_l) = \eta_l(e_l) + \epsilon l(\mu) e_l \int_{E_h}d\eta_h(e)$, 
$\eta_l'(1) = \eta_l(1) -\epsilon l(\mu)\int_{E_h} \frac{e-e_l}{e-1}d\eta_h(e)$.
By choosing $\epsilon>0$ small enough, we get that $(\eta_h',\eta_l')\in \mathcal{K}_E$, and 
     \[ \mu\int_Ed\eta'_h(e)+ \int_{E}\left( \frac{1}{e}-\mu\right) d\eta'_l(e) - \mu\int_Ed\eta_h(e)- \int_{E}\left( \frac{1}{e}-\mu\right) d\eta_l(e)\]
     \[= (1-\mu) \epsilon l(\mu) \int_{E_h} \frac{e-e_l}{e-1}d\eta_h(e)>0\] \end{proof}

A consequence of claim 6 is that,  if  $(\eta_h,\eta_l) \in \argmax_{K_0^E}\{\mu\int_Ed\widetilde{\eta}_h(e)+ \int_{E}\left( \frac{1}{e}-\mu\right) d\widetilde{\eta}_l(e)\}$ and $\supp(\eta_l)\cap E\cap [0,1]\neq\varnothing$ then by there is some  $e_h\in \supp(\eta_h)\cap E\cap [1,\infty]$ such that $\frac{\eta_l(e_h)}{\eta_h(e_h)} < e_hl(\mu)$.

 \begin{claim}
    For any $(\eta_h,\eta_l) \in \argmax_{K_E}\{\mu\int_Ed\widetilde{\eta}_h(e)+ \int_{E}\left( \frac{1}{e}-\mu\right) d\widetilde{\eta}_l(e)\}$. Such that $\supp(\eta_l)\cap E\cap [0,1]\neq\varnothing$. If $\int_E\left( 1 -\frac{1}{e}\right) \eta_l< \int_E \left( 1 -\frac{1}{e}\right)\eta_h $ then for any $e_h \in \supp(\eta_h)\cap E \cap [1,\infty]$, $\frac{\eta_l(e_h)}{\eta_h(e_h)} =e_hl(\mu)$. 
 \end{claim}
 
 \begin{proof}
     Note that $\supp(\eta_l)\cap E\cap [0,1]\neq\varnothing$ implies that there is some $e_l\in E\cap [0,1]$ such that $\eta_h(e_l)>\eta_l(e_l)>0$. In particular, $\eta_l(1) > 0$.
    For contradiction let $\frac{\eta_l(e_h)}{\eta_h(e_l)} < el(\mu)$.
     Construct $\eta_h',\eta_l'$ such that for all $e\not\in \{e_h,e_l,1,0\}$ we have $\eta_\theta'(e) = \eta_\theta(e)$. 
     Define $\eta_h'(e_h) = \eta_h(e_h) -\epsilon$, 
     $\eta_h'(e_l) = \eta_h(e_l) +\epsilon$,
     $1-\int_Ed\eta_h'(e) = 1-\int_Ed\eta_h(e) -\delta$, 
     $\eta_l'(e_h) = \eta_l(e_h) + p$,
     $\eta_l'(e_l) = \eta_l(e_l) + \kappa - p$,
     $\eta_l'(1) = \eta_l(1) -\kappa$.
     Where $\kappa = p\frac{e_h-e_l}{(1-e_l)e_h}$,
     $ p = \epsilon e_ll(\mu)\frac{(1-e_l)e_h}{e_h-1}$,
     $\delta = \epsilon \frac{e_h-e_l}{e_he_l}$,
     $\epsilon > 0$.
     By choosing $\epsilon$ close enough to $0$ we get that $(\eta_h',\eta_l')\in \mathcal{K}_E$, and 
     \[ \mu\int_Ed\eta'_h(e)+ \int_{E}\left( \frac{1}{e}-\mu\right) d\eta'_l(e) - \mu\int_Ed\eta_h(e)- \int_{E}\left( \frac{1}{e}-\mu\right) d\eta_l(e)\]
     \[= (1-\mu)\kappa>0\]\end{proof}

In particular for $(\eta_h,\eta_l) \in \argmax_{K^E_1}\{\mu\int_Ed\widetilde{\eta}_h(e)+ \int_{E}\left( \frac{1}{e}-\mu\right) d\widetilde{\eta}_l(e)\}$, such that $\supp(\eta_l)\cap E\cap [0,1]\neq\varnothing$
and $ \frac{\eta_l(e)}{\eta_h(e)}< el(\mu)$ for some $e\in E\cap [1,\infty]$, it must be that $\int_E\left( 1 -\frac{1}{e}\right) \eta_l= \int_E \left( 1 -\frac{1}{e}\right)\eta_h$.

\begin{claim}
    For any $(\eta_h,\eta_l) \in \argmax_{K^E_1}\{\mu\int_Ed\widetilde{\eta}_h(e)+ \int_{E}\left( \frac{1}{e}-\mu\right) d\widetilde{\eta}_l(e)\}$. Whenever $e_l \in \supp(\eta_h)\cap E \cap [0,1]$ then $\eta_l(e_l) = e_l l(\mu)$.
\end{claim}
\begin{proof}
    Assume for contradiction there is some $e_l<1$ such that $\eta_l(e_l) < e_l l(\mu)\eta_h(e_l)$. As $(\eta_h,\eta_l) \in K^E_1$ we get that $1-\int_{E\cup \{1\}}\frac{1}{e}d\eta_l(e)> 0 $.  
    Construct $\eta_h',\eta_l'$ such that $\eta_h' = \eta_h$ and 
    $\eta'_l(e) = \eta_l(e) $ for all $e\not\in \{e_h,e_l\}$.
    Let $\eta_l'(e_h) =\eta_l(e_h) -\epsilon$, $\eta_l'(e_l) =\eta_l(e_l) +\epsilon$,
    $1-\int_{E\cup \{1\}}\frac{1}{e}d\eta'_l(e) = 1-\int_{E\cup \{1\}}\frac{1}{e}d\eta_l(e) - \epsilon\left(\frac{1}{e_l}-\frac{1}{e_h}\right)$.
    By choosing $\epsilon>0$ small enough we get that $(\eta'_h,\eta'_l)\in \mathcal{K}_1^E$ and 
    \[ \mu\int_Ed\eta'_h(e)+ \int_{E}\left( \frac{1}{e}-\mu\right) d\eta'_l(e) - \mu\int_Ed\eta_h(e)- \int_{E}\left( \frac{1}{e}-\mu\right) d\eta_l(e)\]
     \[= (1-\mu)\epsilon\left(\frac{1}{e_l}-\frac{1}{e_h}\right) >0\]
     Thus the claim follows from contradiction. 
\end{proof}

\begin{claim}
    For any $(\eta_h,\eta_l) \in \argmax_{K_0^E}\{\mu\int_Ed\widetilde{\eta}_h(e)+ \int_{E}\left( \frac{1}{e}-\mu\right) d\widetilde{\eta}_l(e)\}$. If  
 $|\supp(\eta_l)\cap E \cap [0,1]| > 1$, then $\frac{\eta_l(e)}{\eta_h(e)}= l(\mu) e$ for all $e\in \supp(\eta_h)\cap E \cap [0,1]$.
\end{claim}
\begin{proof}
    If  $(\eta_h,\eta_l) \in \argmax_{K_0^E}\{\mu\int_Ed\widetilde{\eta}_h(e)+ \int_{E}\left( \frac{1}{e}-\mu\right) d\widetilde{\eta}_l(e)\}$ then by claim 6 there is some  $e_h\in \supp(\eta_h)\cap E\cap [1,\infty]$ such that $\frac{\eta_l(e_h)}{\eta_h(e_h)} < e_hl(\mu)$. Let $e'<e_l \in \supp(\eta_h)\cap E \cap [0,1] $, for contradiction assume that \\
    Case I: $\eta_l(e')< e'l(\mu)\eta_h(e')$. Construct $(\eta'_h,\eta_l')$ such that $\eta_\theta'(e) = \eta_\theta(e) $ for all $e\not \in \{e_h,e_l,e',1,0\}$. Let $\eta_h'(e_h) = \eta_h(e_h)-\epsilon$, $\eta_h'(e') = \eta_h(e')-\delta$, $\eta_h'(e_l) = \eta_h(e_l)+\epsilon+\delta$,  $\eta_l'(e_h) = \eta_l(e_h)+ q$, $\eta_l'(e') = \eta_l(e')$, $\eta_l'(e_l) = \eta_l(e_l)+\kappa -q$, $\eta_l'(1) = \eta_l(1) -\kappa$. Where $\epsilon < \frac{e_l-e'}{e_h-e_l}\frac{e_h}{e'}\delta$, $\kappa = \frac{q}{e_h} + l(\mu)(\epsilon+ \delta)$, $q = \frac{1-e_l}{e_h-1}(\epsilon+ \delta) e_hl(\mu)$, and $\delta>0$. Choosing $\epsilon$ close to the upper bound and $\delta$ close to $0$ gives that $(\eta_h',\eta_l')\in K^E_0$. 
    \[ \mu\int_Ed\eta'_h(e)+ \int_{E}\left( \frac{1}{e}-\mu\right) d\eta'_l(e) - \mu\int_Ed\eta_h(e)- \int_{E}\left( \frac{1}{e}-\mu\right) d\eta_l(e)\]
     \[= (1-\mu)\kappa> 0 \]
      Case II: $\eta_l(e_l)< e_ll(\mu)\eta_h(e_l)$. Construct $(\eta'_h,\eta_l')$ such that $\eta_\theta'(e) = \eta_\theta(e) $ for all $e\not \in \{e_h,e_l,e',1,0\}$. Let $\eta_h'(e_h) = \eta_h(e_h)+\epsilon$, $\eta_h'(e_l) = \eta_h(e_l)-\delta-\epsilon$, $\eta_h'(e') = \eta_h(e')+\delta$,  $\eta_l'(e_h) = \eta_l(e_h)+ \kappa -\delta e'l(\mu)$, $\eta_l'(e') = \eta_l(e') + \delta e'l(\mu)$, $\eta_l'(e_l) = \eta_l(e_l)$, $\eta_l'(1) = \eta_l(1) -\kappa$. Where $\epsilon >\max\{ \frac{e_l-e'}{e_h-e_l}\frac{e_h}{e'}\delta, \frac{1-e_l}{e_h}\delta\}$, $\kappa = \delta l(\mu)\frac{e_h-e_l}{e_h-1}$, and $\delta>0$. Choosing $\epsilon$ close to the upper bound and $\delta$ close to $0$ gives that $(\eta_h',\eta_l')\in K^E_0$. 
    \[ \mu\int_Ed\eta'_h(e)+ \int_{E}\left( \frac{1}{e}-\mu\right) d\eta'_l(e) - \mu\int_Ed\eta_h(e)- \int_{E}\left( \frac{1}{e}-\mu\right) d\eta_l(e)\]
     \[= (1-\mu)\kappa> 0 \]
\end{proof}

Now we are ready to prove the main result needed for the proof of theorem 3:

\begin{prop}
    $ \argmax_{K_E}\{\mu\int_Ed\widetilde{\eta}_h(e)+ \int_{E}\left( \frac{1}{e}-\mu\right) d\widetilde{\eta}_l(e)\} \subset \textbf{cvx}(\mathcal{T}(\mathcal{K}_E))$.
\end{prop}
\begin{proof}
    As mentioned before, I prove this result in two cases

\textbf{Case I:} $(\eta_h,\eta_l) \in argmax_{K_E}\{\mu\int_Ed\widetilde{\eta}_h(e)+ \int_{E}\left( \frac{1}{e}-\mu\right) d\widetilde{\eta}_l(e)\}$ such that $\supp(\eta_l)\cap E\cap [0,1]\neq\varnothing$ and $\frac{\eta_l(e)}{\eta_h(e)} = el(\mu)$ for all $e\in \supp(\eta_h)\cap E\cap [1,\infty]$. 

In this case, by proof of claim 7, it must be that $(\eta_h,\eta_l)  \in K_1^E$. Let $W_I$ be the set of $(\eta_h,\eta_l)$ that satisfy the conditions of case I. Claims 10 -12 establish this case.

\begin{claim}
    
    For some $(\eta_h,\eta_l) \in \argmax_{K^E_1}\{\mu\int_Ed\widetilde{\eta}_h(e)+ \int_{E}\left( \frac{1}{e}-\mu\right) d\widetilde{\eta}_l(e)\}$. If $\supp(\eta_l)\cap E\cap [0,1]\neq\varnothing$, then $\frac{\eta_l(e)}{\eta_h(e)}\leq1$ for all  $e\in E$. 
\end{claim}

\begin{proof}
    By claim 4 either $\supp(\eta_l)\cap E\cap [\frac{1}{l(\mu)},\infty] = \varnothing$ or $|\supp(\eta_l)\cap E\cap [1,\infty]| = 1$. 

    If $\supp(\eta_l)\cap E\cap [1,\infty] = \{e_h\}$ then by $\int_E\left(1-\frac{1}{e}\right)(\eta_h(e)-d\eta_l(e))\geq 0$ we get that $\frac{\eta_l(e_h)}{\eta_h(e_h)}\leq1$.

    If $\supp(\eta_h)\cap E\cap [\frac{1}{l(\mu)},\infty] = \varnothing$ then the claim follows from noting $\frac{\eta_l(e)}{\eta_h(e)}\leq el(\mu)$ for all $e\in E$. \end{proof}

Define 
\[\widetilde{K}^E_1:=\{(\eta_h,\eta_l)\in W_I| ~|supp(\eta_h)\cap E \cap [0,1] | \leq 1 \text{ and } \frac{\eta_h(e)}{\eta_l(e)} = el(\mu)\text{ for all } e\in \supp(\eta_h)\cap E\cap[1,\infty]\}\]

\begin{claim}
    For any $E$ we have $\argmax_{W_I}\{\mu\int_Ed\widetilde{\eta}_h(e)+ \int_{E}\left( \frac{1}{e}-\mu\right) d\widetilde{\eta}_l(e)\} \subset \textbf{cvx}(\widetilde{K}^E_1)$. 
\end{claim}
\begin{proof}
Fix $(\eta_h,\eta_l) \in \argmax_{W_I}\{\mu\int_Ed\widetilde{\eta}_h(e)+ \int_{E}\left( \frac{1}{e}-\mu\right) d\widetilde{\eta}_l(e)$\}. Assume that $|\supp(\eta_h)\cap E\cap [0,1]|> 2$. By claim 8 we get that $\eta_l(e)  = el(\mu) \eta_h(e)$ for all $e\in \supp(\eta_h)\cap E\cap [0,1]$. 

By assumption and claim 5 we know that for any $e>1\in E$, $\eta_h(e)>0 \implies \frac{\eta_l(e)}{\eta_h(e)} = el(\mu)$. Pick some $e_l\in \supp(\eta_h)\cap E\cap [0,1]$ Now construct the following $(\eta_h',\eta_l')$ and $(\eta_h'',\eta_l'')$:

    \[\eta_{\theta}'(e) =\frac{\alpha \eta_{\theta}(e)}{\alpha\int_{E\cap[1,\infty]}d\eta_h(e)+\eta_h(e_l)}~\text{ for all } e\in E\cap [1,\infty]\]

     \[\eta_{\theta}'(e_l) =\frac{ \eta_{\theta}(e_l)}{\alpha\int_{E\cap[1,\infty]}d\eta_h(e)+\eta_h(e_l)}\]

      \[\eta_{\theta}''(e) =\frac{(1-\alpha) \eta_{\theta}(e)}{(1-\alpha)\int_{E\cap[1,\infty]}d\eta_h(e)+\int_{E\cap[0,1]\setminus \{e_l\}}d\eta_h(e)}~\text{ for all } e\in E\cap [1,\infty]\]

       \[\eta_{\theta}''(e) =\frac{ \eta_{\theta}(e)}{(1-\alpha)\int_{E\cap[1,\infty]}d\eta_h(e)+\int_{E\cap[0,1]\setminus \{e_l\}}d\eta_h(e)}~\text{ for all } e\in E\cap[0,1]\setminus \{e_l\}\]
       Note that $\frac{\int_{E\cap[1,\infty]}d\eta_l(e)}{\int_{E\cap[1,\infty]}d\eta_h(e)}\leq 1$ as $W_I\subset K^E_1$, this implies $\frac{\int_{E\cap[1,\infty]}d\eta'_l(e)}{\int_{E\cap[1,\infty]}d\eta'_h(e)}\leq 1$ and $\frac{\int_{E\cap[1,\infty]}d\eta''_l(e)}{\int_{E\cap[1,\infty]}d\eta''_h(e)}\leq 1$. Thus to show that  $(\eta'_h,\eta'_l),~(\eta''_h,\eta''_l)\in W_I$. We need to verify the following sets of inequalities can be satisfied simultaneously:
       
       \[\frac{\left(\frac{1}{e_l}-1\right)\eta_l(e_l)}{\int_{E\cap[1,\infty]}\left(1-\frac{1}{e}\right)d\eta_l(e)}< \alpha< \frac{\int_{E\cap[1,\infty]}\left(1-\frac{1}{e}\right)d\eta_l(e)+\int_{E\cap[0,1]\setminus \{e_l\}}\left(1-\frac{1}{e}\right)d\eta_l(e)}{\int_{E\cap[1,\infty]}\left(1-\frac{1}{e}\right)d\eta_l(e)}\]
       
        \[\frac{\left(\frac{1}{e_l}-1\right)(\eta_h(e_l)-\eta_l(e_l))}{\int_{E\cap[1,\infty]}\left(1-\frac{1}{e}\right)(d\eta_h(e)-d\eta_l(e))}\leq \alpha
        \]
        
        \[\alpha\leq \frac{\int_{E\cap[1,\infty]}\left(1-\frac{1}{e}\right)(d\eta_h(e)-d\eta_l(e))+\int_{E\cap[0,1]\setminus \{e_l\}}\left(1-\frac{1}{e}\right)(d\eta_h(e)-d\eta_l(e))}{\int_{E\cap[1,\infty]}\left(1-\frac{1}{e}\right)(d\eta_h(e)-d\eta_l(e))}\]
        
        The first two inequalities are satisfied simultaneously for some $\alpha$ as $(\eta_h,\eta_l)\in K_1^E$. Similarly, the third and fourth inequalities also hold simultaneously. Finally, 
        note that the third inequality implies the first one as 
        \[\frac{\left(\frac{1}{e_l}-1\right)\eta_l(e_l)}{\int_{E\cap[1,\infty]}\left(1-\frac{1}{e}\right)d\eta_l(e)} < \frac{\left(\frac{1}{e_l}-1\right)(\eta_h(e_l)-\eta_l(e_l))}{\int_{E\cap[1,\infty]}\left(1-\frac{1}{e}\right)(d\eta_h(e)-d\eta_l(e))}\]
        iff
        \[ \left(\frac{1}{e_l}-1\right)\eta_l(e_l)\int_{E\cap[1,\infty]}\left(1-\frac{1}{e}\right)d\eta_h(e) < \left(\frac{1}{e_l}-1\right)\eta_h(e_l)\int_{E\cap[1,\infty]}\left(1-\frac{1}{e}\right)d\eta_l(e)\]
        which follows from 
        \[\int_{E\cap [1,\infty]}\left(1-\frac{1}{e}\right)\left[ \frac{\eta_l(e_l)}{\eta_h(e_l)}-\frac{\eta_l(e)}{\eta_h(e)}\right]d\eta_h(e)\]
        \[= l(\mu)\int_{E\cap [1,\infty]}\left(1-\frac{1}{e}\right)\left[ e_l-e\right]d\eta_h(e) < 0\]
        Also, the fourth inequality implies the second one as 
        \[\int_{E\cap[0,1]\setminus \{e_l\}}\left(1-\frac{1}{e}\right)d\eta_l(e) \int_{E\cap[1,\infty]}\left(1-\frac{1}{e}\right)d\eta_h(e)\]
        \[> \int_{E\cap[0,1]\setminus \{e_l\}}\left(1-\frac{1}{e}\right)d\eta_h(e) \int_{E\cap[1,\infty]}\left(1-\frac{1}{e}\right)d\eta_l(e) \]
        This follows as 
        \[\int_{E\cap[0,1]\setminus \{e_l\}} \left(\frac{1}{e'}-1\right)\left[\int_{E\cap [1,\infty]}\left(1-\frac{1}{e}\right)\left[ \frac{\eta_l(e')}{\eta_h(e')}-\frac{\eta_l(e)}{\eta_h(e)}\right]d\eta_h(e)\right] d\eta_h(e')\]
        \[= l(\mu)\int_{E\cap[0,1]\setminus \{e_l\}} \left(\frac{1}{e'}-1\right)\left[\int_{E\cap [1,\infty]}\left(1-\frac{1}{e}\right)\left[ e'-e\right]d\eta_h(e)\right] d\eta_h(e') < 0\]
        
    Thus we can always choose $\alpha \in [0,1]$ that satisfies the requirements. Thus we get that $(\eta'_h,\eta'_l) \in \widetilde{K}_1^E$ and $(\eta''_h,\eta''_l) \in W_I$. Recursively applying this argument proves the claim.
\end{proof}

Let $\mathcal{B}(\mathcal{K}_E):= \{(\eta_h,\eta_l)\in \mathcal{K}_E|~|\supp(\eta_h)\cap E|\leq 2\}$

\begin{claim}
    $\widetilde{K}_1^E \subset \textbf{cvx}(\mathcal{B}(\mathcal{K}_E))$.
\end{claim}

\begin{proof}

    If $\supp(\eta_h)\cap [0,1]=\varnothing$ then $(\eta_h,\eta_l) = \sum_{e\in E}\eta_h(e)(\eta^e_h,\eta^e_l)$. Where $\eta^e_l(e)=\frac{\eta_l(e)}{\eta_h(e)}$, $\eta^e_l(1)=1-\frac{\eta_l(e)}{\eta_h(e)}$  and $\eta^e_h(e) = 1$ for all $e\in E$.

    If  $\supp(\eta_h)\cap [0,1] = \{e_l\}$. The claim follows when $|\supp(\eta_h)\cap [1,\infty]|=1$. Assume that $|\supp(\eta_h)\cap [1,\infty]|>1$, also for any $e>1\in E$, $\eta_h(e)>0 \implies \frac{\eta_l(e)}{\eta_h(e)} = el(\mu)$. Pick some $e_h \in \supp(\eta_h)\cap [1,\infty]$.  Now construct the following $(\eta_h',\eta_l')$ and $(\eta_h'',\eta_l'')$:
    
   
    \[\eta_{\theta}'(e_h) =\frac{ d\eta_{\theta}(e_h)}{\alpha\eta_h(e_l)+\eta_h(e_h)}\]

     \[\eta_{\theta}'(e_l) =\frac{ \alpha d\eta_{\theta}(e_l)}{{\alpha\eta_h(e_l)+\eta_h(e_)h)}} \]

        \[\eta_{\theta}''(e) =\frac{ d\eta_{\theta}(e)}{(1-\alpha)\eta_h(e_l)+\int_{E\cap[1,\infty]\setminus\{e_h\}}d\eta_h(e)}~\text{ for all } e\in E\cap[1,\infty]\setminus\{e_h\}\]

      \[\eta_{\theta}''(e_l) =\frac{(1-\alpha) \eta(e_l)}{(1-\alpha)\eta_h(e_l)+\int_{E\cap[1,\infty]\setminus\{e_h\}}d\eta_h(e)}\]

    By claim 10 we know that $\frac{\eta_l(e)}{\eta_h(e)}\leq 1$ for all $e\in E$. Thus to show that  $(\eta'_h,\eta'_l),~(\eta''_h,\eta''_l)\in \widetilde{K}_1^E$. We need to verify the following sets of inequalities can be satisfied simultaneously:
       \[\frac{\left(1-\frac{1}{e_h}\right)\eta_l(e_h)}{\left(\frac{1}{e_l}-1\right)\eta_l(e_l)}> \alpha > \frac{\left(\frac{1}{e_l}-1\right)\eta_l(e_l)+\int_{E\cap[1,\infty]\setminus\{e_h\}}\left(\frac{1}{e}-1\right)d\eta_l(e)}{\left(\frac{1}{e_l}-1\right)\eta_l(e_l)}\]
       
        \[\frac{\left(1-\frac{1}{e_h}\right)(\eta_h(e_h)-d\eta_l(e_h))} {
        \left ( \frac{1}{e_l}-1 \right ) (\eta_h(e_l)-\eta_l(e_l))}\geq \alpha
        \]
        
        \[\alpha\geq \frac{\int_{E\cap[1,\infty]\setminus\{e_h\}}\left(\frac{1}{e}-1\right)(d\eta_h(e)-d\eta_l(e))+\left ( \frac{1}{e_l}-1 \right ) (\eta_h(e_l)-\eta_l(e_l))}{
        \left ( \frac{1}{e_l}-1 \right ) (\eta_h(e_l)-\eta_l(e_l))}\]   
    
    Note that as $(\eta_h,\eta_l)\in \widetilde{K}_1^E$, the first and second inequalities hold simultaneously for some $\alpha$, also the third and fourth inequalities hold simultaneously. The claim follows from noting that the third inequality implies the first one and the fourth inequality implies the second one. 
    Third implies the first as 
    \[\frac{\eta_l(e_h)}{\eta_h(e_h)}= l(\mu)e_h > l(\mu)e_l = \frac{\eta_l(e_l)}{\eta_h(e_l)}\]
    and the fourth inequality implies the second one as 
    \[\frac{\int_{E\cap[1,\infty]\setminus\{e_h\}}\left(1-\frac{1}{e}\right)(d\eta_l(e))}{\int_{E\cap[1,\infty]\setminus\{e_h\}}\left(1-\frac{1}{e}\right)(d\eta_h(e))} \geq l(\mu) > l(\mu)e_l = \frac{\eta_l(e_l)}{\eta_h(e_l)} \]

     Thus we can always choose $\alpha \in [0,1]$ that satisfies the requirements. Thus we get that $(\eta'_h,\eta'_l) \in \mathcal{B}(\mathcal{K}_E)$ and $(\eta''_h,\eta''_l) \in \widetilde{K}_1^E$. Recursively applying this argument proves the claim.
\end{proof}
The proof of the first case then follows from noting that by definition $\mathcal{B}(\mathcal{K}_E)\subset \mathcal{T}(\mathcal{K}_E)$.\\

\textbf{Case II:} $(\eta_h,\eta_l) \in argmax_{K_E}\{\mu\int_Ed\widetilde{\eta}_h(e)+ \int_{E}\left( \frac{1}{e}-\mu\right) d\widetilde{\eta}_l(e)\}$ such that $\supp(\eta_l)\cap E\cap [0,1]\neq\varnothing$ and $\frac{\eta_l(e)}{\eta_h(e)} < el(\mu)$ for some $e\in \supp(\eta_h)\cap E\cap [1,\infty]$. 

In this case, by proof of claims 6 and 7, it must be that $\int_E\left(1-\frac{1}{e}\right)d\eta_h(e) = \int_E\left(1-\frac{1}{e}\right)d\eta_l(e)$. 

\begin{claim}
     For any $(\eta_h,\eta_l) \in \argmax_{K_E}\{\mu\int_Ed\widetilde{\eta}_h(e)+ \int_{E}\left( \frac{1}{e}-\mu\right) d\widetilde{\eta}_l(e)\}$ such that $\supp(\eta_l)\cap E\cap [0,1] \neq \varnothing$. If there exists some $e\in \supp(\eta_h)\cap E\cap [1,\infty]$ such that $\frac{\eta_l(e)}{\eta_h(e)}<el(\mu)$ then  $|\supp(\eta_l)\cap E\cap [1,\infty]| = 1$.
\end{claim}
\begin{proof}
   Assume that there exists $e_4<1\in E$ such that $\eta_l(e_4)> 0 $. Assume for contradiction that $|\supp(\eta_l)\cap E\cap [1,\infty]| > 1$, then there exist $e_2, e_3 \in \supp(\eta_l)\cap E \cap [1,\infty]$ where $e_2> e_3$. By claim 5 we get that $\frac{\eta_l(e_2)}{\eta_h(e_2)} = e_2l(\mu)$ and $\frac{\eta_l(e_3)}{\eta_h(e_3)} = e_3l(\mu)$. By assumption and claim 5 we get that there exists $e_1 \in \supp(\eta_h)\cap E \cap [1,\infty]$ such that $e_1>e_2$ and $\eta_l(e_1) =0$.
   
   Construct $(\eta'_h,\eta_l')$ such that $\eta_\theta'(e) = \eta_\theta(e)$ for all $e\not\in \{e_1,e_2,e_3,e_4,1\}$.
   Let $\eta'_h(e_1) = \eta_h(e_1) -q$, $\eta'_h(e_2) = \eta_h(e_2) -\epsilon$, 
   $\eta'_h(e_4) = \eta_h(e_4) -\delta$, $\eta'_h(e_3) = \eta_h(e_3) +q+ \epsilon+ \delta$, 
   $\eta'_l(e_1) = 0$, 
   $\eta'_l(e_2) = \eta_l(e_2) -\epsilon e_2 l(\mu)$, 
   $\eta'_l(e_4) = \eta_l(e_4) -\delta e_4 l(\mu)$, 
   $\eta'_l(e_3) = \eta_l(e_3) +(q+ \epsilon+ \delta)e_3 l(\mu)$,
   and $\eta'_l(1) = \eta_l(1) - ql(\mu)$.
   Where $\epsilon = \kappa q \frac{\frac{e_1-e_3}{e_1}+ \frac{e_3-1}{e_4}}{(e_2-e_3)\left(\frac{1}{e_4}-\frac{1}{e_2}\right)}$, $\delta = q\frac{1}{e_3-e_4}\left[ \frac{\kappa\frac{e_1-e_3}{e_1}+ \frac{e_3-1}{e_2} +\frac{e_3-1}{e_4}(\kappa -1)}{\frac{1}{e_4}-\frac{1}{e_2}} \right]$. By choosing some $\kappa> 1$ and $q>0$ small enough we get that $(\eta_h',\eta_l') \in \mathcal{K}_E$. The claim follows by noting that
    \[ \mu\int_Ed\eta'_h(e)+ \int_{E}\left( \frac{1}{e}-\mu\right) d\eta'_l(e) - \mu\int_Ed\eta_h(e)- \int_{E}\left( \frac{1}{e}-\mu\right) d\eta_l(e)\]
     \[=(1-\mu) q l(\mu)>0\]
\end{proof}

\begin{claim}
    For any $(\eta_h,\eta_l) \in \argmax_{\mathcal{K}_E}\{\mu\int_Ed\widetilde{\eta}_h(e)+ \int_{E}\left( \frac{1}{e}-\mu\right) d\widetilde{\eta}_l(e)\}$. If  
 $|\supp(\eta_h\cap E \cap [1,\infty]| = 1$, then $|\supp(\eta_l)\cap E \cap [0,1]| = 1$.
\end{claim}
\begin{proof}
    Let $(\eta_h,\eta_l) \in \argmax_{\mathcal{K}_E}\{\mu\int_Ed\widetilde{\eta}_h(e)+ \int_{E}\left( \frac{1}{e}-\mu\right) d\widetilde{\eta}_l(e)\}$, such that  
$\supp(\eta_h\cap E \cap [1,\infty] = \{e_h\}$ and $\frac{\eta_l(e_h)}{\eta_h(e_h)} < e_hl(\mu)$. Assume for contradiction $|\supp(\eta_l)\cap E \cap [0,1]| > 1$, then by claim 13 we get that $\frac{\eta_l(e)}{\eta_h(e)}=el(\mu)$ for all $e\in \supp(\eta_h)\cap E\cap [0,1]$. By proof of claim 6 and 7 we conclude that $\int_E\left(1-\frac{1}{e}\right) d\eta_h(e) = \int_E\left(1-\frac{1}{e}\right) d\eta_l(e)$. Also by corollary 1 we get that $\eta_h(e_h) = 1-\int_{E\cap [0,1]} d\eta_h(e)$. Combining these we get that 
\[ \eta_l(e_h)  = \eta_h(e_h) - \frac{e_h}{e_h-1}\int_{E\cap [0,1]} \left(\frac{1}{e}-1\right)(1-el(\mu))d\eta_h(e)\]
The feasibility constraint can then be simplified as
\[ 1- \int_{E\cap [0,1]} \left (1 + \frac{e_h(1-e)}{e(e_h-1)}\right) d\eta_h(e)\geq 0\]
The above defines a half space for the feasible choice of $\eta_h$. As the $\eta_l$ is determined completely by the choice of $\eta_h$, the objective function can be expressed as a linear functional of $\eta_h$ alone. In particular, the extreme points of the feasible region are candidate solutions for the optimal $\eta_h$. These extreme points are such that either $\eta_h(e_h) =1$ or $\eta_h(e) = \frac{e(e_h-1)}{e_h-e}$ for some $e\in E\cap [0,1]$ and zero for all other $e'\in E\cap [0,1]$. In particular notice that at optimum $|\supp(\eta_l)\cap E \cap [0,1]| = 1$. 

\end{proof}

Define $\widetilde{K}_0^E$ as the set of all $(\eta_h,\eta_l) \in \mathcal{K}_E$ such that  $|\supp(\eta_h) \cap E \cap [1,\infty]| \leq 2$ and $ \exists~e\in \supp(\eta_h) \cap E \cap [1,\infty] \text{ such that } \frac{\eta_l(e)}{\eta_h(e)} < el(\mu) $

\begin{claim}
    For any $(\eta_h,\eta_l) \in \argmax_{\mathcal{K}_E}\{\mu\int_Ed\widetilde{\eta}_h(e)+ \int_{E}\left( \frac{1}{e}-\mu\right) d\widetilde{\eta}_l(e)\}$ such that $\supp(\eta_l)\cap E\cap [0,1] \neq \varnothing$. If there exists some $e\in \supp(\eta_h)\cap E\cap [1,\infty]$ such that $\frac{\eta_l(e)}{\eta_h(e)}<el(\mu)$ then $(\eta_h,\eta_l)\in \textbf{cvx}(\widetilde{K}^E_0)$.
\end{claim}
\begin{proof}
     Define $E_0 = \{e\in E\cap [1,\infty] ~|~e\in \supp(\eta_h)\setminus \supp(\eta_l)\}.$ If $|E_0|=1$, then the claim follows from claim 13. Assume that $|E_0|>1$. By claim 13 we get that $|\supp(\eta_l)\cap E\cap [1,\infty]| = 1$.
     
     Pick some $e_h \in E_0$.  Now construct the following $(\eta_h',\eta_l')$ and $(\eta_h'',\eta_l'')$:
    
   
    \[\eta_{\theta}'(e_h) =\frac{ \eta_{\theta}(e_h)}{\alpha\int_{E\setminus E_0}d\eta_h(e)+\eta_h(e_h)}\]

     \[\eta_{\theta}'(e) =\frac{ \alpha \eta_{\theta}(e)}{{\alpha\int_{E\setminus E_0}d\eta_h(e)+\eta_h(e_h)}}~\text{ for all } e\in E\setminus E_0 \]
     
 \[\eta_{\theta}''(e) =\frac{ \eta_{\theta}(e)}{(1-\alpha)\int_{E\setminus E_0}d\eta_h(e)+\int_{E_0\setminus\{e_h\}}d\eta_h(e)}~\text{ for all } e\in E_0\setminus \{e_h\}\]

    \[\eta_{\theta}''(e) =\frac{ (1-\alpha) \eta_{\theta}(e)}{(1-\alpha)\int_{E\setminus E_0}d\eta_h(e)+\int_{E_0\setminus\{e_h\}}d\eta_h(e)}~\text{ for all } e\in E\setminus E_0\]

    When $\alpha = \frac {\left( 1- \frac{1}{e_h}\right)\eta_h(e_h)}{\int_{E\setminus E_0}\left( \frac{1}{e} -1\right)(d\eta_h(e)-d\eta_l(e))}$, we get that $ \int_{E\setminus (E_0\setminus\{e_h\})}\left( 1- \frac{1}{e}\right)(d\eta'_h(e)-d\eta'_l(e) ) = \int_{E\setminus e_h}\left( 1- \frac{1}{e}\right)(d\eta''_h(e)-d\eta''_l(e) ) =0$. Moreover, by construction $ \int_{E\setminus (E_0\setminus\{e_h\})}\left( 1- \frac{1}{e}\right)d\eta'_l(e) , \int_{E\setminus e_h}\left( 1- \frac{1}{e}\right)d\eta''_l(e)  \geq 0$. Finally $(\eta_h',\eta_l')\in \mathcal{K}_E$ and $(\eta_h'',\eta_l'')\in \mathcal{K}_E$ follows from noting the following:
    \[\int_{E\setminus E_0} d\eta_h(e) -d\eta_l(e)\geq \int_{E\setminus E_0}(1-el(\mu)) d\eta_h(e)\]
    \[ > \int_{E\setminus E_0} (1-e) d\eta_h = \int_{E\setminus E_0} \frac{l(\mu) e}{l(\mu)}\left( \frac{1}{e}-1\right)d\eta_h(e)\]
    \[ = \frac{1}{l(\mu)}\int_{E\setminus E_0}\left( \frac{1}{e}-1\right) d \eta_l(e) \geq 0\]
    \[\implies \int_E d\eta'_l(e),~\int_E\eta''_l(e) < 1\]
 
\end{proof}

Define
\[\mathcal{T}(\mathcal{K}_E):= \{(\eta_h,\eta_l)\in \mathcal{K}_E|~|\supp(\eta_h)\cap E|\leq 3, |\supp(\eta_l)\cap E|\leq 2\}\]

\begin{claim}
    $ \argmax_{\widetilde{K}_0^E}\{\mu\int_Ed\widetilde{\eta}_h(e)+ \int_{E}\left( \frac{1}{e}-\mu\right) d\widetilde{\eta}_l(e)\} \subset \mathcal{T}(\mathcal{K}_E)$. 
\end{claim}
\begin{proof}
    Fix some $(\eta^*_h,\eta^*_l) \in   \argmax_{\widetilde{K}_0^E}\{\mu\int_Ed\widetilde{\eta}_h(e)+ \int_{E}\left( \frac{1}{e}-\mu\right) d\widetilde{\eta}_l(e)\}$. Assume that $|\supp(\eta^*_l)\cap E\cap [0,1]| \geq 2$ otherwise the claim is trivial.
    When $|\supp(\eta^*_h)\cap E\cap [1,\infty]| <2$, the claim follows from claim 14. Assume that $\supp(\eta_h)\cap E\cap [1,\infty] =\{\overline{e},e_h\}$, where $e_h<\overline{e}$. We get by claim 5, claim 8, and claim 9 that $\eta^*_l(e) = el(\mu)$ for all $e\in E\setminus\{\overline{e}\}$ and  $\eta^*_l(\overline{e}) = 0$. Thus we get the following: 
    \[ \eta_h(e_h) \geq \frac{1}{e_h-1}\int_{E\cap [0,1]}(1-e)d\eta_h(e)\]
    \[ \eta_h(\overline{e}) = 1-\int_{E\setminus \overline{e}}d\eta_h(e)\]
    and 
    \[ \int_E\left(1-\frac{1}{e}\right)d\eta_h(e) = \int_E\left(1-\frac{1}{e}\right)d\eta_l(e)\]
    
    \[\implies  \eta_h(e_h) = \frac{1-\frac{1}{\overline{e}}+ \int_{E\cap [0,1]} \left( l(\mu)(1-e) - \left( \frac{1}{e}-\frac{1}{\overline{e}}\right) \right)d\eta_h(e)} {\frac{1}{e_h} - \frac{1}{\overline{e}} + l(\mu) (e_h -1)}\]
    and 
    \[\implies \eta_h(\overline{e}) = 1- \frac{1-\frac{1}{\overline{e}}}{\frac{1}{e_h} - \frac{1}{\overline{e}} + l(\mu) (e_h -1)} - \int_{E\cap [0,1]} \frac{l(\mu)(e_h-e) - \left( \frac{1}{e}-\frac{1}{_he}\right)}{\frac{1}{e_h} - \frac{1}{\overline{e}} + l(\mu) (e_h -1)}\]
    In particular, the feasibility requires 
    \[ \frac{1}{e_h} - \frac{1}{\overline{e}} + l(\mu)(e_h-1)\overset{(1)}{\geq}  1 - \frac{1}{\overline{e}} + \int_{E\cap [0,1]} \left( l(\mu)(1-e) - \left( \frac{1}{e}-\frac{1}{\overline{e}}\right) \right)d\eta_h(e) \overset{(2)}{\geq}  0 \]

    \[ \left( l(\mu) -\frac{1}{e_h}\right) (e_h -1) \overset{(3)}{\geq} \int_{E\cap [0,1]} \left( l(\mu) - \frac{1}{ee_h}\right) (e_h-e) d\eta_h(e) \]

    \[ 1 \overset{(4)}{\geq}  \frac{1}{(\overline{e}-1)(e_h -1)} \int_{E\cap [0,1]} \left( \frac{(1-e)(\overline{e}-e_h)}{e_h} + \frac{(\overline{e}-e)(e_h-1)}{e}\right) d\eta_h(e)\]

    Note that $ (4) \implies (2)$, also $(2)~\&~(3)\implies (1)$, also note that $(4)$ simplifies to $ 1\geq  \int_{E\cap [0,1]} \frac{e_h-e}{e_h-1} \left( \frac{\overline{e}(e_h-1) + e(\overline{e}-e_h)}{e_he(\overline{e}-1)}\right) d\eta_h(e)$ and that $ \frac{e_h-e}{e_h-1} \left( \frac{\overline{e}(e_h-1) + e(\overline{e}-e_h)}{e_he(\overline{e}-1)}\right)>1$ for all $e\in E\cap[0,1]$.
    
    \textbf{If $ l(\mu) \geq \frac{1}{e_h}$} then  $(4)\implies(3)$. This follows from:

    \[\frac{ l(\mu) -\frac{1}{ee_h}}{l(\mu)-\frac{1}{e_h}}\leq \frac{\overline{e}(e_h-1) + e(\overline{e}-e_h)}{e_he(\overline{e}-1)} \text{ for all } e\in E\cap [0,1]\]
    
    Denote the first coordinate of $\widetilde{K}_0^E$ by $J_0^E$. Thus the optimization problem can be relaxed as 
    \[\max_{\eta_h\in J_0^E} T(\eta_h)\]
    \[\text{such that } G(\eta_h) \geq 0\]
    Where $G(\eta_h) \geq 0$ defines a closed convex half space as $G(\eta_h) =  1 -  \int_{E\cap [0,1]} \frac{e_h-e}{e_h-1} \left( \frac{\overline{e}(e_h-1) + e(\overline{e}-e_h)}{e_he(\overline{e}-1)}\right) d\eta_h(e)$.
    The feasible region $G(\eta_h) \geq 0 $ is closed and convex with extreme points given by $\eta_h$ such that $\int_{E\cap [0,1]}d\eta_h(e) = 0 $ or $\eta_h(e) = \frac{e_h-1}{e_h-e} \left( \frac{e_he(\overline{e}-1)}{\overline{e}(e_h-1) + e(\overline{e}-e_h)}\right)$ for some $e\in E\cap [0,1]$ and $\eta_h(e') = 0$ for all $e' \in E\cap [0,1] \setminus \{e\}$.

    \textbf{If $ l(\mu) < \frac{1}{e_h}$} then $(3)$ is given by 
    \[ 1\leq \int_{E\cap [0,1]} \frac{ l(\mu) -\frac{1}{ee_h}}{l(\mu)-\frac{1}{e_h}}\frac{e_h-e}{e_h-1}d\eta_h(e)\]
    Thus the optimization problem is given by 
    \[\max_{\eta_h\in J_0^E} T(\eta_h)\]
    \[\text{such that } G(\eta_h) \geq 0\]
    \[ H(\eta_h) \geq 0 \]
    Where $H(\eta_h) = \int_{E\cap [0,1]} \frac{ l(\mu) -\frac{1}{ee_h}}{l(\mu)-\frac{1}{e_h}}\frac{e_h-e}{e_h-1}d\eta_h(e) -1 $. The halfspaces defined by $(3)$ and $(4)$ have none empty intersection as 
    \[\frac{ l(\mu) -\frac{1}{ee_h}}{l(\mu)-\frac{1}{e_h}}>\frac{\overline{e}(e_h-1) + e(\overline{e}-e_h)}{e_he(\overline{e}-1)} \text{ for all } e\in E\cap [0,1]\]
   Thus, the feasible region $\{\eta_h | G(\eta_h) \geq 0 \geq -H(\eta_h)\}$ is closed, nonempty and convex with extreme points given by $\eta_h$ such that $\eta_h(e) = \frac{e_h-1}{e_h-e} \left( \frac{e_he(\overline{e}-1)}{\overline{e}(e_h-1) + e(\overline{e}-e_h)}\right)$ or $\eta_h(e) =  \frac{l(\mu)-\frac{1}{e_h}}{ l(\mu) -\frac{1}{ee_h}}\frac{e_h-1}{e_h-e}$ for some $e\in E\cap [0,1]$ and $\eta_h(e') = 0$ for $e' \in E\cap [0,1] \setminus \{e\}$. Here the first type of extreme points are on the hyperplane $G(\eta_h) =0$ and the second type lies on the hyperplane $H(\eta_h)= 0$. 

    The objective 
    \[T(\eta_h) = \mu\int_Ed\eta_h(e)  + \int_{E\setminus \{\overline{e}\}}\left(\frac{1}{e}-\mu\right) el(\mu)d\eta_h(e)\]
    \[ = \text{constant} + \int_{E\cap [0,1]} l(\mu) \left( (1-\mu_h)\frac{l(\mu)(1-e) - \frac{1}{e}+\frac{1}{\overline{e}}}{\frac{1}{e_h}-\frac{1}{\overline{e}}+ l(\mu)(e_h-1)}+ (1-e\mu)\right)d\eta_h(e)\]
    This is linear in $\eta_h$ thus the optimum is achieved at an extreme point, thus we prove the claim as for any extreme point it holds that $|\supp(\eta_h)\cap E\cap [0,1] |\leq 1$.

\end{proof}
Claims 10 -12 establish case I and claim 13 - 16 establish case II. Thus proposition 3 follows by combining these with claims 2 and 3. 
\end{proof}

Theorem 3 follows from proposition 3, by constructing the menu such that $\sigma_h(.|h)=\eta_h(.)$ and $\sigma_h(.|h)=\eta_l(.)$ with payments according to proposition 2.

\subsection{Lemma 6: Characterization of Optimal Binary Support Menus}
\begin{lem}
     If $E=\{e_h,e_l\}$, where $e_l<1$ and $e_h>1$. Then $(\sigma_h,\sigma_l)\in \mathcal{K}_E$ if and only if it has the following form:\\
     If $e_h\leq \min\{\frac{1}{l(\mu)},\frac{1}{\mu}\}$ and $\frac{1}{\mu + (1-\mu)l(\mu)e_l} \geq e_h$
     \[\begin{smallmatrix} \sigma_h\\\\
\begin{pmatrix}
\frac{e_h(1-e_l)}{e_h-e_l}\frac{1-l(\mu)e_l}{1-l(\mu)e_he_l} & \frac{e_l(e_h-1)}{e_h-e_l}\frac{1-l(\mu)e_h}{1-l(\mu)e_he_l}& 0\\

\frac{1-e_l}{e_h-e_l}\frac{1-l(\mu)e_l}{1-l(\mu)e_he_l} & \frac{e_h-1}{e_h-e_l}\frac{1-l(\mu)e_h}{1-l(\mu)e_he_l}& \frac{l(\mu)(e_h+e_l-e_he_l-1)}{1-l(\mu)e_he_l}\\

\end{pmatrix} \end{smallmatrix}\]
\[\begin{smallmatrix} \sigma_l\\\\
\begin{pmatrix}
e_hl(\mu)\frac{e_h(1-e_l)}{e_h-e_l}\frac{1-l(\mu)e_l}{1-l(\mu)e_he_l} & e_ll(\mu)\frac{e_l(e_h-1)}{e_h-e_l}\frac{1-l(\mu)e_h}{1-l(\mu)e_he_l}& 1-l(\mu)\left[e_l-e_h(1-e_l)\frac{1-l(\mu)e_l}{1-l(\mu)e_he_l}\right]& 0\\

e_hl(\mu)\frac{1-e_l}{e_h-e_l}\frac{1-l(\mu)e_l}{1-l(\mu)e_he_l} & e_ll(\mu)\frac{e_h-1}{e_h-e_l}\frac{1-l(\mu)e_h}{1-l(\mu)e_he_l}& 1-l(\mu)\left[e_l-e_h(1-e_l)\frac{1-l(\mu)e_l}{1-l(\mu)e_he_l}\right] & l(\mu)\frac{(e_h-1)(1-e_l)}{1-l(\mu)e_he_l}\\

\end{pmatrix} \end{smallmatrix}\]

     If $e_h> \frac{1}{\mu}$ or if  $\frac{1}{l(\mu)}< e_h\leq \frac{1}{\mu}$ and $\frac{1}{e_h}\leq \mu + (1-\mu)e_ll(\mu)$ or if $e_h\leq \min\{\frac{1}{l(\mu)},\frac{1}{\mu}\}$ and $\frac{1}{\mu + (1-\mu)l(\mu)e_l} < e_h$
\[\begin{smallmatrix} \sigma_h\\\\
\begin{pmatrix}
\frac{e_h(1-e_l)}{e_h-e_l} & \frac{e_l(e_h-1)}{e_h-e_l}\\

\frac{1-e_l}{e_h-e_l} & \frac{e_h-1}{e_h-e_l}\\

\end{pmatrix} \end{smallmatrix} ~~~~~
\begin{smallmatrix} \sigma_l\\\\
\begin{pmatrix}
e_ll(\mu)\frac{e_h(1-e_l)}{e_h-e_l} & e_ll(\mu)\frac{e_l(e_h-1)}{e_h-e_l} & 1-l(\mu)e_l\\

e_ll(\mu)\frac{1-e_l}{e_h-e_l} & e_ll(\mu)\frac{e_h-1}{e_h-e_l} & 1-l(\mu)e_l\\

\end{pmatrix} \end{smallmatrix}\]

     and, if  $\frac{1}{l(\mu)} < e_h\leq \frac{1}{\mu}$ and $\frac{1}{e_h}> \mu + (1-\mu)e_ll(\mu)$
     \[\begin{smallmatrix} \sigma_h = \sigma_l\\\\
\begin{pmatrix}
1 & 0\\
\frac{1}{e_h} & 1-\frac{1}{e_h}

\end{pmatrix} \end{smallmatrix} \]
      
\end{lem}
\begin{proof}
     A binary menu wrt $E=\{e_h (>1),e_l (<1)\}$ is of the form the following form for some $x\in (0,1]$.
     \[\sigma_h\begin{smallmatrix} 
\begin{pmatrix} 

x & 1-x &0\\
\frac{x}{e_h} & \frac{1-x}{e_l} & 1 - \frac{x}{e_h} -\frac{1-x}{e_l}\\
 
\end{pmatrix} \end{smallmatrix} ~~~~~\]
\[ \sigma_l
\begin{smallmatrix}
\begin{pmatrix}

\alpha x & \beta(1-x) & 1 -\alpha x -\beta (1-x)& 0\\
\alpha \frac{x}{e_h} & \beta\frac{1-x}{e_l} & 1 -\alpha x -\beta (1-x)& \alpha\left(1-\frac{1}{e_h}\right)x + \beta\left(1-\frac{1}{e_l}\right)(1-x)\\
\end{pmatrix} \end{smallmatrix}\]
Thus the optimization problem in Proposition 2 is simplified to   
 
  \[ \max ~~\mu + \left( \frac{1}{e_h}-\mu\right)\alpha x +  \left( \frac{1}{e_l}-\mu\right)\beta(1-x)\]
   subject to\\
  1)  $x\in[0,1],~~$ 
2)  $\frac{1}{e_h}x+\frac{1}{e_l}(1-x)\leq 1,~~$ 
3) $\alpha x\left(1-\frac{1}{e_h}\right) + \beta(1-x)\left(1-\frac{1}{e_l}\right)\geq 0,~~$ 
4) $\alpha x+\beta(1-x)\leq 1,~~$
5) $\alpha x\left(1-\frac{1}{e_h}\right) + \beta(1-x)\left(1-\frac{1}{e_l}\right)\geq 0,~~$ 
6)  $\alpha\leq \min\{e_hl(\mu), 1\}$ and 
7) $\beta \leq l(\mu)e_l$

Note that 2) $\frac{1}{e_h}x + \frac{1}{e_l}(1-x)\leq 1$ implies $x\geq \frac{e_h(1-e_l)}{e_h-e_l}$. 3) $\alpha x\left(1-\frac{1}{e_h}\right) + \beta(1-x)\left(1-\frac{1}{e_l}\right)\geq 0$ implies $\beta \leq \alpha\frac{e_l(e_h-1)}{e_h(1-e_l)}\frac{x}{1-x}$. 4) $\alpha x+\beta(1-x)\leq 1,~~$ implies that $\alpha \leq 1-\frac{e_h(1-e_l)}{e_l(e_h-1)}\frac{1-x}{x}(1-\beta)$.\\

If $\frac{1}{e_h}< \mu$ then the revenue is decreasing in $\alpha$ and $x$. Setting $x= \frac{e_h(1-e_l)}{e_h-e_l}$, we get that $\alpha = \beta$ by 3) and 4). Thus the revenue is $\mu + \left( \frac{1}{e_l}-\mu - \frac{e_h(1-e_l)}{e_h-e_l}\left( \frac{1}{e_l}-\frac{1}{e_h}\right)\right)\beta = \mu + \left( \frac{1}{e_l}-\mu - \frac{1-e_l}{e_l}\right)\beta$. Thus revenue is increasing in $\beta$. Hence the optimal experiment is such that $\sigma_h(e_h)= \frac{e_h(1-e_l)}{e_h-e_l}$ and $\alpha = \beta = e_ll(\mu)$.  

If $l(\mu)>\frac{1}{e_h}> \mu$ the revenue is increasing in $\alpha$. Thsu or fixed $x,\beta$ we have $\alpha = \min \{ e_hl(\mu), 1-\frac{e_h(1-e_l)}{e_l(e_h-1)}\frac{1-x}{x}(1-\beta)\}$. As $e_hl(\mu) )> 1$ we get that $\alpha =  1-\frac{e_h(1-e_l)}{e_l(e_h-1)}\frac{1-x}{x}(1-\beta)$ Then the revenue is given by:
\[\mu +  (1-\frac{e_h(1-e_l)}{e_l(e_h-1)}\frac{1-x}{x}(1-\beta))x \left( \frac{1}{e_h}-\mu\right)+ \left( \frac{1}{e_l}-\mu\right)\beta(1-x)\]
\[ = \mu - \frac{e_h(1-e_l)}{e_l(e_h-1)} \left( \frac{1}{e_h}-\mu\right) + \beta  \left( \frac{1}{e_l}-\mu\right) + x\left[\left( \frac{1}{e_h}-\mu\right)(1+\frac{e_h(1-e_l)}{e_l(e_h-1)}(1-\beta)) - \beta \left( \frac{1}{e_l}-\mu\right) \right] \]

\[ = \mu - \frac{e_h(1-e_l)}{e_l(e_h-1)} \left( \frac{1}{e_h}-\mu\right) + \beta  \left( \frac{1}{e_l}-\mu\right)\]
\[ + x\left[\left( \frac{1}{e_h}-\mu\right)\left(1+\frac{e_h(1-e_l)}{e_l(e_h-1)}\right) - \beta \left( \frac{1}{e_l}-\mu + \frac{e_h(1-e_l)}{e_l(e_h-1)}\left( \frac{1}{e_h}-\mu\right)\right) \right] \]

\[= \mu - \frac{e_h(1-e_l)}{e_l(e_h-1)} \left( \frac{1}{e_h}-\mu\right) + \beta  \left( \frac{1}{e_l}-\mu\right) + x \frac{e_h-e_l}{e_l(e_h-1)}\left[ \frac{1}{e_h}-\mu - \beta(1-\mu)\right]\]
Whenever $\frac{1}{e_h}-\mu -\beta(1-\mu) > 0 $ then the revenue is increasing in $x$, thus the optimal experiment is such that $x=1$ and $\alpha = 1$. Whenever $\frac{1}{e_h}-\mu -\beta(1-\mu)<0 $ then revenue is decreasing in $x$ for fixed $\beta$.  Thus $x= \frac{e_h(1-e_l)}{e_h-e_l}$, this gives us that the optimal experiment is such that $\sigma_h(e_h)= \frac{e_h(1-e_l)}{e_h-e_l}$ and $\alpha = \beta = e_ll(\mu)$.\\

If $e_h\leq \frac{1}{\mu},\frac{1}{l(\mu)}$, the revenue is increasing in $\beta$ and $\alpha$. For fixed $x,\alpha$, we get that $\beta = \min\{e_l(l(\mu), ~\alpha\frac{e_l(e_h-1)}{e_h(1-e_l)}\frac{x}{1-x}\}$. 
If $\beta = \alpha\frac{e_l(e_h-1)}{e_h(1-e_l)}\frac{x}{1-x}$, then the revenue is given by 
\[ \mu + x\alpha\left[\frac{1}{e_h}-\mu + \left(\frac{1}{e_l}-\mu\right) \frac{e_l(e_h-1)}{e_h(1-e_l)}\right]\]
\[= \mu + \alpha x\frac{e_h-e_l}{e_h(1-e_l)}(1-\mu)\]
The expression is increasing in $x$. As $\alpha\frac{e_l(e_h-1)}{e_h(1-e_l)}\frac{x}{1-x} <  e_ll(\mu)$ we get that 
\[x\leq \frac{l(\mu)e_h(1-e_l)}{\alpha(e_h-1)+l(\mu)e_h(1-e_l))}\]
As the revenue is increasing in both $\alpha$ and $x$, thus for fixed $\alpha$ set $x =  \frac{l(\mu)e_h(1-e_l)}{\alpha(e_h-1)+l(\mu)e_h(1-e_l)}$. Then the revenue is given by 
\[ \mu + (1-\mu)l(\mu)\alpha\frac{e_h-e_l}{\alpha(e_h-1)+l(\mu)e_h(1-e_l)}\]
This expression is increasing in $\alpha$. As $x\geq \frac{e_h(1-e_l)}{e_h-e_l}$ we get that 
\[\frac{l(\mu)e_h(1-e_l)}{\alpha(e_h-1)+l(\mu)e_h(1-e_l)} \geq  \frac{e_h(1-e_l)}{e_h-e_l}\]
\[ \implies l(\mu)(e_h-e_l) \geq \alpha(e_h-1)+l(\mu)e_h(1-e_l)\]
\[\implies \alpha \leq l(\mu)e_l \]
Thus the optimal experiment has $x= \frac{e_h(1-e_l)}{e_h-e_l}$ and $\alpha=\beta= e_ll(\mu)$.\\ 
If $\beta = e_ll(\mu) $ then the revenue is increasing in $\alpha$ and $x$. In particular for fixed $x$ we get that $x\alpha \geq (1-x)\frac{e_h(1-e_l)}{ e_l(e_h-1)}$ and $\alpha = \min\{e_hl(\mu), 1-\frac{e_h(1-e_l)}{e_l(e_h-1)}\frac{1-x}{x}(1-e_ll(\mu))\}$. For $x\geq \frac{e_h(1-e_l)}{e_h-e_l}\frac{1-l(\mu)e_l}{1-l(\mu))e_le_h}$, we get that $\alpha = e_hl(\mu)$, in which case the revenue is 
\[\mu + e_ll(\mu)\left(\frac{1}{e_l}-\mu\right) + l(\mu)x\mu(e_l-e_h)\]
this is decreasing in $x$ thus it's optimal to set $x =  \frac{e_h(1-e_l)}{e_h-e_l}\frac{1-l(\mu)e_l}{1-l(\mu))e_le_h}$, $\alpha= e_h l(\mu)$ and $\beta= e_ll(\mu)$.

If $x< \frac{e_h(1-e_l)}{e_h-e_l}\frac{1-l(\mu)e_l}{1-l(\mu))e_le_h}$ we get $\alpha = 1-\frac{e_h(1-e_l)}{e_l(e_h-1)}\frac{1-x}{x}(1-e_ll(\mu))$, then the revenue is given by 
\[= \mu - \frac{e_h(1-e_l)}{e_l(e_h-1)} \left( \frac{1}{e_h}-\mu\right) + \beta  \left( \frac{1}{e_l}-\mu\right) + x \frac{e_h-e_l}{e_l(e_h-1)}\left[ \frac{1}{e_h}-\mu - \beta(1-\mu)\right]\]
Whenever $\frac{1}{e_h}-\mu -\beta(1-\mu) > 0 $ then the revenue is increasing in $x$, thus the optimal experiment is such that $x=\frac{e_h(1-e_l)}{e_h-e_l}\frac{1-l(\mu)e_l}{1-l(\mu))e_le_h}$, $\alpha = e_hl(\mu)$ and $\beta = e_ll(\mu)$. Whenever $\frac{1}{e_h}-\mu -\beta(1-\mu)<0 $ then revenue is decreasing in $x$. Thus $x= \frac{e_h(1-e_l)}{e_h-e_l}$, this gives us that the optimal experiment is such that $\sigma_h(e_h)= \frac{e_h(1-e_l)}{e_h-e_l}$ and $\alpha = \beta = e_ll(\mu)$.

Finally to conclude note that for $e_h\leq \frac{1}{l(\mu)}, \frac{1}{\mu}$ the revenue for $x=\frac{e_h(1-e_l)}{e_h-e_l}\frac{1-l(\mu)e_l}{1-l(\mu))e_le_h}$, $\alpha = e_hl(\mu)$ and $\beta = e_ll(\mu)$ is given  by 
\[\mu + (1-\mu)l(\mu) - \mu l(\mu)\frac{(e_h-1)(1-e_l)}{1-l(\mu)e_he_l} \]
revenue when $x= \frac{e_h(1-e_l)}{e_h-e_l}$ and $\alpha=\beta = e_ll(\mu)$ is given by 
\[\mu + (1-\mu)l(\mu)e_l\]
Note that 
\[\mu + (1-\mu)l(\mu) - \mu l(\mu)\frac{(e_h-1)(1-e_l)}{1-l(\mu)e_he_l}  \geq \mu + (1-\mu)l(\mu)e_l\]
\[\iff (1-\mu)\geq \mu \frac{e_h-1}{1-l(\mu)e_he_l}\]
\[\iff 1 \geq e_h(\mu + (1-\mu)l(\mu)e_l)\]

\end{proof}

\subsection{Lemma 7: Existence of Partial Pooling}

 By theorem 1, theorem 3, and lemma 6 we get that if $m$ is a menu with partial pooling then it must be that $m\in \textbf{cvx}(\mathcal{T}(E))\setminus\textbf{cvx}(\mathcal{B}(E))$.
Thus I will focus on menus in the set $\mathcal{T}(E)\setminus\mathcal{B}(E)$. 
 by proposition 3 we get that i f $m\in \mathcal{M}_E^r\cap [\mathcal{T}(E)\setminus\mathcal{B}(E)]$ then $\supp(\sigma_m(.|h))\cap E =\{\overline{e},e_h,e_l\}$ and $\supp(\sigma_m(.|l))\cap E =\{e_h,e_l\}$ where $\overline{e}>e_h>1>e_l$. Also, $\frac{\sigma_m(e_h|l)}{\sigma_m(e_h|h)}= e_hl(\mu)$. To establish existence I will assume $\min\{l(\mu), \mu + (1-\mu)l(\mu) \frac{e_he(\overline{e}-1)}{\overline{e}(e_h-1) + e(\overline{e}-e_h)} \}\geq \frac{1}{e_h}$. Thus by proof of claim 16, we get the optimal menu has the following form for some $x\in (0,1]$.
     \[\sigma_h\begin{smallmatrix} 
\begin{pmatrix} 

1-x-y & y &x \\
\frac{1-x-y}{\overline{e}} & \frac{y}{e_h} & \frac{x}{e_l}\\
 
\end{pmatrix} \end{smallmatrix} ~~~~~\]
   \[\sigma_l\begin{smallmatrix} 
\begin{pmatrix} 

0 & ye_hl(\mu) &\beta x & 1-ye_hl(\mu) - \beta x\\
0 & yl(\mu) &  \frac{\beta x}{e_l} & 1-ye_hl(\mu) - \beta x \\
 
\end{pmatrix} \end{smallmatrix} ~~~~~\]

In particular we get that $\beta = \frac{e_h-1}{1-e_l}e_l\frac{y}{x}l(\mu)$ and $y = \frac{1-\frac{1}{\overline{e}}-\left(\frac{1}{e_l}-\frac{1}{\overline{e}} \right)x}{\frac{1}{e_h}-\frac{1}{\overline{e}}}$. The revenue is given by 
\[\mu + (1-\mu)l(\mu) y\frac{e_h-e_l}{1-e_l}\]

Note that we need $x+y\leq 1$, $e_hl(\mu)y+ \beta x\leq 1$ and $\beta \leq e_l l(\mu)$. Form $e_hl(\mu)y+ \beta x\leq 1$ we get that 
\[ l(\mu)y\left(e_h+\frac{e_h-1}{1-e_l}e_l\right)\leq  1\]
\[\implies y \leq \frac{1}{l(\mu)}\frac{1-e_l}{e_h-e_l} \]
\[\implies x\geq \frac{1-\frac{1}{\overline{e}}- \frac{1}{l(\mu)}\frac{1-e_l}{e_h-e_l}\left( \frac{1}{e_h}-\frac{1}{\overline{e}}\right)}{\frac{1}{e_l}-\frac{1}{\overline{e}}}\]
From $\beta \leq l(\mu)e_l$ we get that 
\[ 1-\frac{1}{\overline{e}}-\left(\frac{1}{e_l}-\frac{1}{\overline{e}} \right)x \leq x\left(\frac{1}{e_h}-\frac{1}{\overline{e}}\right)\frac{1-e_l}{e_h-1}\]
\[\implies x \geq \frac{ 1-\frac{1}{\overline{e}}}{\left(\frac{1}{e_l}-\frac{1}{\overline{e}} \right)+\left(\frac{1}{e_h}-\frac{1}{\overline{e}}\right)\frac{1-e_l}{e_h-1}}\]
From $x+y\leq 1$ we get that 
\[\frac{1-\frac{1}{\overline{e}}-\left(\frac{1}{e_l}-\frac{1}{e_h} \right)x}{\frac{1}{e_h}-\frac{1}{\overline{e}}}\leq 1\]
\[ \implies x \geq \frac{1-\frac{1}{e_h}}{\frac{1}{e_l}-\frac{1}{e_h}} = \frac{e_l(e_h-1)}{e_h-e_l}\]
The revenue is decreasing in $x$ and the following inequalities hold $\frac{ 1-\frac{1}{\overline{e}}}{\left(\frac{1}{e_l}-\frac{1}{\overline{e}} \right)+\left(\frac{1}{e_h}-\frac{1}{\overline{e}}\right)\frac{1-e_l}{e_h-1}} \leq  \frac{ 1-\frac{1}{\overline{e}}}{\left(\frac{1}{e_l}-\frac{1}{\overline{e}} \right)+\left(\frac{1}{e_h}-\frac{1}{\overline{e}}\right)\frac{1-e_l}{e_h-1}}\leq 1$ and $\frac{1-\frac{1}{\overline{e}}- \frac{1}{l(\mu)}\frac{1-e_l}{e_h-e_l}\left( \frac{1}{e_h}-\frac{1}{\overline{e}}\right)}{\frac{1}{e_l}-\frac{1}{\overline{e}}} \leq 1$. Note that $\frac{ 1-\frac{1}{\overline{e}}}{\left(\frac{1}{e_l}-\frac{1}{\overline{e}} \right)+\left(\frac{1}{e_h}-\frac{1}{\overline{e}}\right)\frac{1-e_l}{e_h-1}} \geq \frac{1-\frac{1}{\overline{e}}- \frac{1}{l(\mu)}\frac{1-e_l}{e_h-e_l}\left( \frac{1}{e_h}-\frac{1}{\overline{e}}\right)}{\frac{1}{e_l}-\frac{1}{\overline{e}}}$ whenever $l(\mu) \leq \frac{e_l(\overline{e}-e_h)+\overline{e}(e_h-1)}{(\overline{e}-1)e_le_h}$. Finally by noting that $\frac{e_l(\overline{e}-e_h)+\overline{e}(e_h-1)}{(\overline{e}-1)e_le_h}\geq 1$ we get that 
 $x = \frac{ 1-\frac{1}{\overline{e}}}{\left(\frac{1}{e_l}-\frac{1}{\overline{e}} \right)+\left(\frac{1}{e_h}-\frac{1}{\overline{e}}\right)\frac{1-e_l}{e_h-1}}$, $y = \frac{ 1-\frac{1}{\overline{e}}}{\left(\frac{1}{e_l}-\frac{1}{\overline{e}} \right)\frac{e_h-1}{1-e_l}+\left(\frac{1}{e_h}-\frac{1}{\overline{e}}\right)}$ and $\beta = l(\mu)e_l$.

\end{sloppypar}
\end{document}